\documentclass[12pt]{article}
\usepackage[top=1in,bottom=1in,left=1in,right=1in]{geometry}
\usepackage{amsmath,amssymb,amsfonts,amsthm}
\usepackage{latexsym}
\usepackage[scr=boondox]{mathalfa}
\usepackage[utf8]{inputenc}
\usepackage{verbatim}
\usepackage{setspace}
\usepackage{dsfont}
\usepackage{color}
\usepackage{tikz}
\usepackage{mathdots}
\usepackage{yhmath}
\usepackage{cancel}
\usepackage{color}
\usepackage{siunitx}
\usepackage{array}
\usepackage{multirow}
\usepackage{amssymb}
\usepackage{gensymb}
\usepackage{tabularx}
\usepackage[normalem]{ulem}
\usepackage{booktabs}
\usetikzlibrary{fadings}
\usetikzlibrary{patterns}
\usetikzlibrary{shadows.blur}
\usetikzlibrary{shapes}
\usepackage{cite}
\usepackage{pifont}
\usepackage{hyperref}
\usepackage{slashed}

\numberwithin{equation}{section}

\newcommand{\p}{\partial}

\newcommand{\bit}{\begin{itemize}}
\newcommand{\eit}{\end{itemize}}
\newcommand{\bd}{\begin{description}}
\newcommand{\ed}{\end{description}}

\newcommand{\bc}{\begin{center}}
\newcommand{\ec}{\end{center}}

\newcommand{\be}{\begin{equation}}
\newcommand{\ee}{\end{equation}}
\newcommand{\bea}{\begin{eqnarray}}
\newcommand{\eea}{\end{eqnarray}}
\newcommand{\bs}{\begin{subequations}}
\newcommand{\es}{\end{subequations}}

\def\p{\partial}

\def\CC{{\mathcal C}}

\def\CH{{\mathcal H}}
\def\CI{{\mathcal I}}
\def\CJ{{\mathcal J}}

\def\CL{{\mathcal L}}
\def\CM{{\mathcal M}}
\def\CN{{\mathcal N}}
\def\CO{{\mathcal O}}

\def\CS{{\mathcal S}}

\def\a{\alpha}

\def\g{\gamma}

\def\th{\theta}

\def\s{\sigma}

\def\G{\Gamma}

\def\Th{\Theta}

\def\0{{(0)}}
\def\1{{(1)}}
\def\2{{(2)}}

\def\bz{{\bar{z}}}

\def\bw{{\bar{w}}}

\def\mem{{\text{mem}}}
\def\shock{{\text{shock}}}

\def\mee{{\mathbb{E}}}
\def\cc{{\text{c.c.}}}
\def\EH{{\text{EH}}}
\def\bdy{{\text{bdy}}}

\def\GI{{\text{GI}}}
\def\VZ{{\text{VZ}}}
\def\KM{{\text{KM}}}

\def\ci{{\mathscr I}}

\begin{document}
\begin{titlepage}
\unitlength = 1mm
\hfill CALT-TH 2024-028
\ \\
\vskip 3cm
\begin{center}

\openup .5em

{\LARGE{An Infrared On-Shell Action and its Implications for Soft Charge Fluctuations in Asymptotically Flat Spacetimes}}

\vspace{0.8cm}
Temple He,$^1$ Ana-Maria Raclariu,$^{2*}$ Kathryn M. Zurek$^1$

\vspace{1cm}

{\it  $^1$Walter Burke Institute for Theoretical Physics \\ California Institute of Technology, Pasadena, CA 91125 USA}\\
{\it  $^2$King's College London, Strand, London, WC2R 2LS}

\vspace{0.8cm}

\begin{abstract}
 
We study the infrared on-shell action of Einstein gravity in asymptotically flat spacetimes, obtaining an effective, gauge-invariant boundary action for memory and shockwave spacetimes. We show that the phase space is in both cases parameterized by the leading soft variables in asymptotically flat spacetimes, thereby extending the equivalence between shockwave and soft commutators to spacetimes with non-vanishing Bondi mass. We then demonstrate that our on-shell action is equal to three quantities studied separately in the literature: $(i)$ the soft supertranslation charge; $(ii)$ the shockwave effective action, or equivalently the modular Hamiltonian;  and $(iii)$ the soft effective action. Finally, we compute the quantum fluctuations in the soft supertranslation charge and, assuming the supertranslation parameter may be promoted to an operator, we obtain an area law, consistent with earlier results showing that the modular Hamiltonian has such fluctuations.
 \end{abstract}

\vspace{1.0cm}
\end{center}
\vspace{30pt}
*Corresponding author: ana-maria.raclariu@kcl.ac.uk
\end{titlepage}
\pagestyle{empty}
\pagestyle{plain}
\pagenumbering{arabic}

\tableofcontents

\section{Introduction}

The goal of this work is to further explore the relation between shockwave spacetimes and asymptotically flat spacetimes (AFSs) highlighted in \cite{He:2023qha}. It was shown therein that a certain class of shockwave metrics is related by a diffeomorphism to asymptotically flat metrics exhibiting the leading gravitational memory effect, which we refer to as memory metrics. This suggests that one should be able to reframe the computation of fluctuations of causal diamonds in a shockwave spacetime \cite{Verlinde:2022hhs} as one in an asymptotically flat spacetime with leading gravitational memory. A key step in this endeavor is to identify the simplest physical observable that captures information about such fluctuations. Since physical observables are by definition invariant under diffeomorphisms, computing them using either the shockwave or the memory metric should yield the same answer. 

An important ingredient in \cite{Verlinde:2022hhs} was an effective action living on the boundary of the causal development of a subregion in flat space, which was first studied in \cite{Verlinde:1991iu}. The shockwave effective action was shown in \cite{Verlinde:2022hhs} to be equivalent to the modular Hamiltonian of the spacetime subregion, assuming that in a quantum theory of gravity, the boost parameter is promoted to an operator living on the bifurcate horizon (in a spirit similar to \cite{Liu:2021kay}). Upon solving the Einstein equation, the result is bilinear in the 't Hooft shockwave variables \cite{tHooft:1996rdg}, and the effective boundary action may be used to compute the two-point function of its fluctuations. The fluctuations in this quantum modular Hamiltonian were then argued to induce large fluctuations in the spacetime subregion and its causal development. 

Sufficiently large spacetime subregions may be well-approximated by an asymptotically flat spacetime. One may then expect spacetime fluctuations to be related to fluctuations in certain asymptotic charge aspects. A first step towards making this intuition precise is to identify a gauge-invariant boundary action that would then describe the dynamics of either soft or shockwave degrees of freedom depending on the choice of gauge.\footnote{By choice of gauge, we mean the choice of fall-off conditions for the gravitational potential at large-$r$. This is the gravitational analog of working in Coulomb versus temporal gauge in Yang-Mills, e.g., see \cite{Ball:2018prg}.} The main result of this paper is the construction of such an action for the leading soft sector of gravity in asymptotically flat spacetimes. The construction provides a new entry in the holographic dictionary for asymptotically flat spacetimes, namely a relationship between an on-shell action with boundary conditions that render the variational principle well-defined and the generating function for correlation functions of currents in a codimension-2 conformal field theory (CFT). Such an action has been previously constructed using effective field theory techniques for a codimension-2 theory of Goldstone modes in \cite{Kapec:2021eug}, and remarkably, we find that our bulk, on-shell computation yields the same action!\footnote{An analogous result was obtained in the context of Abelian gauge theories in \cite{He:2024ddb}.}

It has been long known that boundary terms may need to be added to the Einstein-Hilbert (EH) action for the variational principle in gravity to be well-defined. Hence, the on-shell action is typically non-vanishing and reduces to the Gibbons-Hawking-York (GHY)  boundary term \cite{York, Gibbons-Hawking}. In standard examples of AdS/CFT, one typically imposes Dirichlet boundary conditions, which amounts to fixing the induced metric at the boundary. In this case, the boundary action is proportional to an integral of the trace of the extrinsic curvature. Applying the same boundary conditions in asymptotically flat spacetimes, the on-shell action is found to be proportional to an integral of the Bondi mass aspect \cite{Fabbrichesi:1993kz}. Nevertheless, fixing the induced metric on the boundary is too restrictive, as it disallows flux or gravitational radiation through $\ci^{\pm}$. One may then wonder under what conditions gravitational theories have both gravitational flux and a well-defined variational principle. 

In this paper, we will answer this question for a simple class of spacetimes where only the leading memory mode of the gravitational field is turned on \cite{Strominger:2014pwa}. 
Such spacetimes are equipped with metrics parameterized by a shear given by 
\be 
\label{eq:shear-intro}
	C_{zz}(u,z,\bar{z}) = \p_{z}^2 N(z,\bz) \theta(u - u_0)  - 2 \p_z^2 C^{\rm}(z, \bar{z}) ,
\ee
where $N$ is the memory mode (an excitation on a fixed vacuum giving rise to gravitational memory, so named because of the unit step function $\th(u-u_0)$ describing its effect), while $C$ is the Goldstone mode parametrizing the moduli space of asymptotically flat vacua \cite{Strominger:2013jfa, Kapec:2021eug, Kapec:2022hih, Capone:2023roc}. The modes $C,N$ jointly parameterize the leading soft sector of the gravitational phase space \cite{Strominger:2017zoo, Donnay:2018neh, Freidel:2022skz}, and such metrics were shown in \cite{He:2023qha} to be related via a linearized diffeomorphism to shockwave metrics of the form studied in \cite{Verlinde:2022hhs}, characterized by an $\mathcal{O}(1)$ potential localized in null time in the large-$r$ limit. It was shown in \cite{He:2023qha} that for the case of a vanishing Bondi mass aspect, such diffeomorphisms preserve the symplectic form. Furthermore, an identification between $C,N$ and the shockwave momentum $P_u$ and the null shift $X^u$ was proposed, namely
\begin{align}\label{eq:intro-PX}
	P_u(z,\bz) = \frac{1}{32\pi G_N} \Box^2 N , \qquad X^u(z,\bz) = - C(z,\bz) .
\end{align}
In the first part of this paper, we generalize this identification to spacetimes with a non-trivial Bondi mass aspect. In particular, we show that diffeomorphisms linear in $N$ continue to preserve the symplectic form, and hence the phase space is still parameterized by the same variables given in \eqref{eq:intro-PX}.

Furthermore, by imposing the boundary condition $\delta \p_z^2 C = \delta \p_\bz^2 C = 0$, we search for a boundary action to be added to the EH action that both ensures the variational principle is satisfied and is also invariant under the time-dependent diffeomorphisms relating the memory and shockwave metrics (see \eqref{eq:full-diffeo}). 
The boundary action is derived from phase space exact terms in the symplectic potential at $\ci^{\pm}$ and therefore transforms under field-dependent diffeomorphisms that are non-vanishing near the boundary \cite{Donnelly:2016auv}. Demanding this action to be invariant under diffeomorphisms preserving the boundary conditions, independently at $\ci^{+}$ and $\ci^-$, requires the further addition of a boundary term bilinear in $C_{zz}$ and $N_{zz}.$  For the profile \eqref{eq:shear-intro}, this localizes to a codimension-2 surface also known as a corner,\footnote{For empty finite causal diamonds, this is a bifurcate horizon.} and we obtain the on-shell action in the soft sector\footnote{Despite being written as an integral over $\mathscr{I}^+_-$, the action \eqref{eq:gi-bdry-intro} includes contributions from both $\mathscr{I}^{\pm}$.} 
\be 
\label{eq:gi-bdry-intro}
	S_{\bdy}^{\GI} = - \frac{1}{8\pi G_N} \int_{\ci^+_-} d^2z\,\p_z^2 C \p_\bz^2 N  + \mathcal{O}(N^2).
\ee
The action \eqref{eq:gi-bdry-intro} actually includes an additional term proportional to the Bondi mass aspect, and the $uu$-component of the vacuum Einstein equation at $\ci^+$ further allows us to trade the Bondi mass aspect for a term quadratic in $N$ with a divergent coefficient. However, since our analysis is linear in $N$, this term does not explicitly appear in \eqref{eq:gi-bdry-intro}.  By construction, this action is found to agree with the boundary action of a shockwave solution to the vacuum Einstein equation. In this case, $C_{zz}$ contains no memory mode $N$; rather, $N$ has transmuted into parametrizing the leading component of the shockwave potential at large-$r$. This results in the corner term vanishing, and we again obtain \eqref{eq:gi-bdry-intro} as expected.

 We would like to emphasize that the variational principle is in general \textit{not} well-defined for asymptotically flat spacetimes with arbitrary flux \cite{Ruzziconi:2020wrb}. The main assumption leading to a well-defined variational principle here is the restriction to shears given by the simple form \eqref{eq:shear-intro}. Equivalently, we are considering radiation that only consists of leading memory modes and employing the special boundary condition that fixes the zero mode $C$ (thereby also fixing the supertranslation degree of freedom).

Given the identification \eqref{eq:intro-PX}, the gauge-invariant action \eqref{eq:gi-bdry-intro} can be seen to agree with the shockwave effective action studied in \cite{Verlinde:2022hhs}, which is
\begin{align}
	S_\VZ = \int_{S^2} d^2z\, X^u P_u,
\end{align}
where $S^2$ is the bifurcate horizon on the causal diamond and is analogous to $\ci^+_-$. Consequently, a main outcome of our analysis is that it provides a derivation of the effective action considered in \cite{Verlinde:2022hhs} as an on-shell action associated with a shockwave spacetime, or equivalently, an on-shell action for the Goldstone and memory modes of asymptotically Minkowski spacetimes.
Interestingly, to linear order in $N$, \eqref{eq:gi-bdry-intro} agrees not only with the effective action studied in \cite{Verlinde:2022hhs}, but also with both the soft supertranslation charge and the soft effective action found in \cite{Kapec:2021eug}. This leads to another important result of our work, namely the reinterpretation of the two-point function of fluctuations in the modular Hamiltonian presented in  \cite{Verlinde:2022hhs} in terms of the two-point function of fluctuations in the soft supertranslation charge regarded as a bilinear operator in $C$ and $N$.

The term quadratic in $N$ further appears to schematically reproduce the soft gravitational S-matrix, which exponentiates infrared divergences subject to an identification of the ratio between short and long distance regulators at $\ci^\pm$ with the ratio of ${\rm UV}$ and ${\rm IR}$ regulators employed in the computation of momentum space scattering amplitudes. As such, \eqref{eq:gi-bdry-intro} suggests a deeper relation between the gravitational dressing and the on-shell action in the soft sector of gravity, which would be interesting to be made more precise. 

The paper is organized as follows. In Section~\ref{sec:prelims}, we review two diffeomorphic asymptotic solutions to the Einstein equation: asymptotically flat spacetimes with a shear given by \eqref{eq:shear-intro} and shockwave spacetimes with a gravitational potential $g_{uu}^{(0)} = -\p_z \p_{\bz} N \delta(u - u_0).$ In Section \ref{sec:thooft}, we generalize the analysis in \cite{He:2023qha} to the case where the Bondi mass aspect is non-vanishing. In Section \ref{sec:bdry-act}, we discuss the relation between the on-shell actions associated with the memory and shockwave spacetimes. Finally, in Section~\ref{sec:equiv-actions} we show that the gauge-invariant on-shell action associated with these spacetimes agrees to linear order in $N$ with the soft supertranslation charge, the Verlinde-Zurek effective action \cite{Verlinde:2022hhs}, and the soft effective action \cite{Kapec:2021eug}.\footnote{This effective action also includes an $\mathcal{O}(N^2)$ term that reproduces the exponent of the soft S-matrix of Weinberg \cite{Weinberg:1965aa}. Our analysis could be generalized to include terms quadratic in $N$, in which case we expect to also recover this term.} We conclude with a brief summary and future directions in Section~\ref{sec:discussion}.

\section{Memory spacetimes and the soft commutator}
\label{sec:prelims}

We begin by considering an asymptotically flat spacetime (AFS) in $(3+1)$-dimensions, whose metric near $\ci^+$ takes the general form \cite{Bondi:1962px,Sachs:1962zza,Barnich:2010eb}
\be\label{eq:afs-met}
\begin{split}
	ds^2 &= \left(-\frac{R_{\gamma}}{2} + \frac{2 m_B(u,z,\bz)}{r}\right)du^2 - 2du \, dr + 2r^2 \gamma_{z\bz} dz\,  d\bz \\
	&\qquad + \left(r C_{zz}(u,z,\bz) dz^2 + D^z C_{zz}(u,z,\bz) du\, dz + \cc \right) + \cdots ,
\end{split}
\ee
where $m_B$ is the Bondi mass aspect, $C_{zz}$ the shear,  $R_{\gamma}$ the curvature of the transverse metric $\gamma_{z\bz}$, $\cc$ denotes the complex conjugate, and $\cdots$ denotes the subleading contributions in a large-$r$ expansion that are determined by the Einstein equation. For simplicity, we will work in planar null coordinates $(u, r, x^{A})$, where $x^A = (z,\bar{z})$, $\gamma_{z\bz} = 1$, and $R_{\gamma} = 0$. With this choice of coordinates, we have
\begin{align}
	C^{\bz\bz} = C_{zz}, \qquad   D^\bz = D_z = \p^\bz = \p_z, \qquad \Box \equiv D^zD_z + D^\bz D_\bz = 2 \p_z\p_\bz .
\end{align}

It will be useful for us to also obtain the subleading terms in the metric. In particular, these will be needed in order to get the correct leading order constraint equation at $\ci^+$, and will contribute non-trivially to the symplectic potential evaluated in \eqref{eq:AF-sp}. Imposing Bondi gauge \cite{Bondi:1962px,Sachs:1962zza,Tamburino:1966zz,Barnich:2010eb}
\begin{align}\label{eq:bondi}
	g_{rr} = g_{rA} = 0 , \qquad \p_r \sqrt{\det \big( r^{-2}g_{AB}\big)} = 0,
\end{align}
and requiring the vacuum Einstein equation to hold order by order in a large-$r$ expansion, we compute in Appendix~\ref{app:AFS} that to first subleading order the AFS metric is given by 
\begin{align}\label{eq:Mink-vacua}
\begin{split}
	ds^2 &= - 2\,du\,dr + 2 r^2\,dz\,d\bz \\
	&\qquad +  \frac{2m_B}{r}  du^2 +  \frac{1}{4r^2} C_{zz}C^{zz}  du\,dr +  C_{zz}C^{zz} \, dz\,d\bz \\
	&\qquad + \big( r C_{zz} \,dz^2 + \p_\bz C_{zz}   du\,dz + \cc \big) + \cdots ,
\end{split}
\end{align}
where, in the vacuum, $m_B$ obeys the constraint equation
\begin{align}\label{eq:mB-constraint}
	\p_u m_B = \frac{1}{4} \big( \p_z^2 N^{zz} + \cc \big) - \frac{1}{4} N_{zz} N^{zz} , \qquad N_{zz}(u,z,\bz) \equiv \p_u C_{zz}(u,z,\bz).
\end{align}
Here, $N_{zz}$ is the news tensor and captures the gravitational radiation leaving $\ci^+$.

For the purposes of this paper, we will focus on ``soft'' or memory shear profiles, which we define as \cite{He:2014laa,Strominger:2014pwa,Donnay:2018neh}
\be \label{eq:soft-shear}
C_{zz}(u,z,\bar{z}) = \p_{z}^2 N(z,\bz) \theta(u - u_0) - 2 \p_z^2 C^{\rm}(z, \bar{z})  ,
\ee
where $\th(u-u_0)$ is the unit step function. Note that $N(z,\bz)$ captures the leading gravitational memory effect and will be referred to as the memory mode \cite{Strominger:2014pwa}, and $C(z,\bz)$ parameterizes the infinite degeneracy of vacua in AFSs and will be referred to as the Goldstone mode \cite{Strominger:2013jfa,Compere:2016hzt,Compere:2016jwb}. From the definition of $N_{zz}$ given in \eqref{eq:mB-constraint}, it is clear that
\begin{align}
	N_{zz}(u,z,\bz) = \p_z^2 N(z,\bz) \delta(u-u_0) .
\end{align}
We will refer to the family of metrics \eqref{eq:Mink-vacua} with the memory shear profile \eqref{eq:soft-shear} as \emph{memory metrics}. In this case, the Bondi mass constraint \eqref{eq:mB-constraint} becomes 
\be 
\label{eq:afs-constraint}
	\p_u  m_B = \frac{1}{2}\p_z^2 \p_{\bz}^2 N \delta(u - u_0) - \frac{1}{4} \delta(u - u_0)^2 \p_z^2 N \p_{\bz}^2 N.
\ee
Note the appearance of a term quadratic in $N$, which corresponds to a gravitational contribution to the stress tensor. The $\delta(u - u_0)^2$ leads to divergences in the change of the Bondi mass aspect along $\ci^+$, as well as in the boundary actions discussed in Section \ref{sec:bdry-act} below. Similar terms were discarded in \cite{tHooft:1987vrq, tHooft:1996rdg}. In Section~\ref{sec:thooft}, we will mostly be interested in a family of metrics obtained from memory metrics of the form \eqref{eq:Mink-vacua} by performing a specific time-dependent diffeomorphism that preserves the vacuum shear profile $C$. For simplicity we will consider diffeomorphisms \textit{linear} in the memory mode $N$, in which case these metrics will be simply related to \eqref{eq:Mink-vacua} via the Lie derivative. As such, we will keep only terms linear in $N$. Nevertheless, the terms quadratic in $N$ may for instance be obtained by demanding that the image of \eqref{eq:Mink-vacua} under the full diffeomorphism obeys the Einstein equation. We will comment on the physical significance and implications of these terms further in Section \ref{sec:discussion}.

Having written down the memory metric explicitly in \eqref{eq:Mink-vacua} with the shear given in \eqref{eq:soft-shear}, we can proceed to compute the gravitational symplectic potential. As reviewed in Appendix \ref{app:symp-pot}, given a variation of the EH action with respect to the metric, so that $g_{\mu\nu} \rightarrow g_{\mu\nu} +  \delta g_{\mu\nu}$, the standard textbook derivation of the symplectic potential is given by (e.g., see \cite{Carroll:2004st})\footnote{The symplectic potential acquires an additional contribution in the presence of matter. This is reviewed in Appendix \ref{app:symp-pot}.} 
\be \label{eq:symp-pot}
\begin{split}
	\Theta(g, \delta g) &= \frac{1}{16\pi G_N} \int_{\p\CM} d\Sigma_\s \big( g^{\sigma\mu} \nabla^\nu \delta g_{\mu\nu} - g^{\mu\nu}\nabla^\s \delta g_{\mu\nu} \big) \equiv \frac{1}{16\pi G_N} \int_{\p\CM} d\Sigma_\s \, \th^\s(g, \delta g) , 
\end{split}
\ee
where $\p\CM$ is the boundary of the spacetime manifold, and we define $\th^\s(g, \delta g)$ to be the integrand.\footnote{Depending on conventions, $\th^\s$ or $-\th^\s$ is referred to as the symplectic potential density.} Here $\delta$ is the exterior derivative on phase space, which implies that the symplectic potential is a one-form on phase space.
Substituting the AFS metric \eqref{eq:Mink-vacua} into \eqref{eq:symp-pot}, we immediately obtain 
\be 
\label{eq:AF-sp}
\begin{split}
	\theta^u &= \mathcal{O}(r^{-3}), \quad \theta^z = \mathcal{O}(r^{-3}),  \\
	\theta^r 
	&= \frac{1}{2 r^2} \left[ 4  \delta m_B - \big(  \p_{\bz}^2 \delta C_{zz} + \cc \big) - \frac{1}{4} \delta\big(C_{zz} N_{\bz\bz} + \cc \big) + \big( \delta C_{zz} N_{\bz\bz} + \cc \big) \right]  + \mathcal{O}(r^{-3}).
\end{split}
\ee
Taking into account the $r^2$ factor from the measure $d\Sigma_\sigma$, at large $r$ it suffices to only consider $\th^r$. Specializing to memory metrics by imposing \eqref{eq:soft-shear}, we obtain
\be
\label{eq:soft-symp} 
\begin{split}
	\theta^r_{\rm mem} &= \frac{1}{2r^2}\bigg[ 4  \delta m_B + 4 \p_z^2 \p_{\bz}^2 \delta C - 2 \p_z^2 \p_{\bz}^2 \delta N \theta(u - u_0) + \frac{1}{4} \delta\left(\p_z^2 N \p_{\bz}^2 N \right) \delta(u - u_0)  \\
	&\qquad + \frac{1}{2}\delta \left( \p_z^2 C \p_{\bz}^2 N + \cc  \right)\delta(u - u_0) - 2 \left(\p_z^2 \delta C \p_{\bz}^2N + \cc   \right) \delta(u - u_0) \bigg] ,
\end{split}
\ee
where we have used the identity $\delta(x) \theta(x) = \frac{1}{2} \delta(x)$. We now evaluate the symplectic potential \eqref{eq:symp-pot} over  $\ci^+$,  which is defined by the $r = \infty$ hyperplane and is hence parametrized by the normal one-form $n_\s = (0,1,0,0)$. It follows that on $\ci^+$
\begin{align}
	\theta^{\s} d\Sigma_\s =  \theta^r \sqrt{-g}\, du\,d^2z,
\end{align}	
and so the symplectic potential for the memory metric is given by
\begin{align}\label{eq:symp-pot-coords}
\begin{split}
	\Th^+_\mem (g, \delta g) &= \frac{1}{16\pi G_N} \int_{\ci^+} du\,d^2z\,\sqrt{-g} \th^r_\mem(g, \delta g) \\
	&= \frac{1}{32\pi G_N} \int_{\ci^+} du\,d^2z  \bigg[ 4 \delta m_B  + \frac{1}{4} \delta \big( \p_z^2 N \p_\bz^2 N  \big) \delta(u-u_0)  \\
	&\qquad + \frac{1}{2} \delta \big(  \p_z^2 C \p_\bz^2 N  + \cc \big)\delta(u-u_0) - 2\big( \p_z^2 \delta C \p_\bz^2 N + \cc \big) \delta(u-u_0)  \bigg].
\end{split}
\end{align}
Here we noted that total derivatives with respect to $z,\bz$ vanish when integrated over the celestial sphere.

We can now straightforwardly compute the symplectic form $\Omega$, which is defined to be
\begin{align}\label{eq:symp-form-def}
\begin{split}
	\Omega \equiv \delta \Theta.
\end{split}
\end{align}
In particular, all phase space exact terms in the symplectic potential will be eliminated since the phase space exterior derivative obeys $\delta^2 = 0$. Therefore, only the last term in \eqref{eq:soft-symp} contributes to the memory symplectic form, yielding 
\be 
\begin{split}
\label{eq:mem-sf}
	\Omega_{\rm mem}^+ =  - \frac{1}{16\pi G_N} \int d^2z \left(\delta \p_{\bz}^2 N  \wedge \delta \p_z^2 C + \cc \right) = - \frac{1}{8\pi G_N} \int d^2z \, \delta\p_\bz^2 N \wedge \delta \p_z^2 C,
\end{split}
\ee
where in the second equality we integrated by parts to combine the two terms. We can now invert the symplectic form to obtain the Poisson bracket\footnote{We fix the sign by recalling the standard symplectic form in Darboux coordinates is $\Omega = \delta p \wedge \delta x$, and this gives rise to the Poisson bracket $\{p, x\} = -1$.} 
\begin{align}
\label{eq:CN}
\begin{split}
	&\big\{ \p_\bz^2 N(z,\bz), \p_w^2 C(w,\bw) \big\} = 8\pi G_N \delta^2(z-w) \\
	\implies\quad & \big\{ N(z,\bz) , C(w,\bw) \big\} = 4 G_N |z-w|^2 \log|z-w|^2 ,
\end{split}
\end{align} 
where we obtained the second line using the fact 
\begin{align}\label{eq:useful}
\begin{split}
	\p_w^2 \big( |z-w|^2\log|z-w|^2 \big) = \frac{\bz - \bw}{z-w}, \qquad \p_\bz \frac{1}{z-w} = 2 \pi \delta^2(z-w).
\end{split}
\end{align}

We can promote the Poisson bracket to a quantum commutator via the identification $[\cdot, \cdot]$ with $i\{\cdot, \cdot\}$, in which case we get the canonical commutation relation of soft variables \cite{He:2014laa}:\footnote{Recall that here $\gamma_{z\bz} = 1$, which differs from the planar coordinate convention adopted in \cite{He:2023qha}, where $\gamma_{z\bz} = 2$. More generally, we have $\{N(z,\bz), C(w,\bw)\} = 4G_{N} \sqrt{\gamma_{z\bz} \gamma_{w\bw}} |z - w|^2 \log|z - w|^2$  \cite{He:2014laa}.}
\be 
\label{eq:soft-comm}
	[N(w,\bw),C(z,\bz)] = 4 i G_N |z - w|^2 \log|z-w|^2\,.
\ee

\section{Shockwave spacetimes with general $m_B$}
\label{sec:thooft}

It was shown in \cite{He:2023qha} for the case $m_B = 0$ that the soft commutator \eqref{eq:soft-comm} may be related via a diffeomorphism to the 't Hooft commutators appearing in shockwave geometries. These are characterized by metrics with a localized gravitational potential term $g_{uu}$ that does not asymptote to $0$ in the limit as $r \to \infty$.  In this section, we generalize that analysis by including metrics with non-zero Bondi mass and show that the commutation relation remains unchanged. We review the diffeomorphism considered in \cite{He:2023qha} in Section~\ref{ssec:diffeo}. We then review how to relate the 't Hooft commutator with the soft commutator when $m_B=0$ in Section~\ref{ssec:mB0}. Finally, we generalize that analysis to arbitrary $m_B$ in Section~\ref{ssec:gen-mB}.

\subsection{From memory to shockwave metric}\label{ssec:diffeo}

Metrics of the form \eqref{eq:Mink-vacua} admit a well-known set of asymptotic symmetries consisting (at least) of supertranslations \cite{Bondi:1962px,Sachs:1962zza,Barnich:2010eb} and superrotations \cite{Barnich:2011mi} (see \cite{Campiglia:2014yka,Freidel:2021fxf} for further enhancements). Establishing these symmetries crucially relies on the boundary conditions usually imposed when analyzing gravity in AFS, namely that in the limit $r \rightarrow \infty$, the metric asymptotes to Minkowski space in one of an infinity of vacua \cite{Strominger:2013jfa}. It was shown in \cite{Geiller:2022vto,Geiller:2024amx} that a relaxation of the fall-off conditions at large-$r$ leads to an enlargement of the allowed diffeomorphisms near $\ci^{\pm}$ that now include time-dependent supertranslations in the $(u,r,z,\bar z)$ basis of the form
\be 
\label{eq:full-diffeo}
	\xi^\mu = \left(F, \xi^{r (0)}  + \frac{1}{r} \xi^{r (1)} + \mathcal{O}(r^{-2}), - \frac{1}{r} \p^A F + \frac{1}{r^2}\xi^{A (2)} + \mathcal{O}(r^{-3}) \right),
\ee
where $F \equiv F(u,z,\bar{z}).$  Both the memory metrics \eqref{eq:Mink-vacua} and the shockwave metrics \eqref{eq:shock-metric} below may then be regarded as members of a larger class of metrics with relaxed fall-off conditions (in particular $\mathcal{O}(1)$ gravitational potentials) as $r \rightarrow \infty$. In the following analysis, we will provide evidence that diffeomorphisms \eqref{eq:full-diffeo} with $F$ given below in \eqref{eq:F-def}, in spite of having finite support at $\ci^+$, relate metrics in the same gauge equivalence class. Consequently, such solutions to the Einstein equation should give rise to the same physical observables. The components $\xi^{r(0)}$ and $\xi^{A(2)}$ are found by imposing the Bondi gauge constraints given in \eqref{eq:bondi} (see \cite{Strominger:2017zoo} for the computation in the special case where $F$ is time-independent),\footnote{We note that unlike \cite{Geiller:2022vto,Geiller:2024amx}, we still demand the determinant condition to be preserved, which can be done consistently with the Einstein equation by turning on $g_{uu}$ and $g_{ur}$ at $\mathcal{O}(1)$, as well as $g_{uz}$ at $\mathcal{O}(r)$, e.g., see also \cite{He:2023qha}.} which gives the constraints
\begin{align}\label{eq:xi-r0}
	\xi^{r(0)} = \frac{1}{2}\Box F = \p_z\p_\bz F, \qquad \xi^{z(2)} = \frac{1}{2} C^{zz} \p_z F,
\end{align}
where $C_{zz}$ is the memory shear \eqref{eq:soft-shear}. Furthermore, we are interested in studying a class of metrics where the memory mode, i.e., the term proportional to $N$ in \eqref{eq:soft-shear}, is entirely eliminated. In this case $C_{zz}(u,z,\bz) \to -2\p_z^2 C(z,\bz)$ only involves the angular components. We will denote the class of such metrics as \emph{shockwave metrics} for reasons that will soon become apparent. In particular, from \eqref{eq:Mink-vacua}, this implies we require
\begin{align}
\begin{split}
	g_{zz}^\mem + \CL_\xi g^\mem_{zz} &= -2 r \p_z^2 C , \qquad g^\mem_{z\bz} + \CL_\xi g^\mem_{z\bz} = r^2 + 2 \p_z^2 C \p_\bz^2 C .
\end{split}
\end{align}
The first condition is satisfied to linear order in $N$ if we fix 
\begin{align}\label{eq:F-def}
	F(u,z,\bz) = \frac{N(z,\bz)}{2}\th(u-u_0),
\end{align}
whereas the second condition is satisfied to linear order in $N$ if we fix
\begin{align}\label{eq:xi-r1}
\begin{split}
	\xi^{r(1)} = - \frac{1}{2} \bigg[ \big( \p_z C_{\bz\bz} \p_{z} F + \cc \big) + \frac{1}{2} \big( \p_z^2 F C_{\bz\bz} + \cc \big)   \bigg] . 
\end{split}
\end{align}
We have now completely fixed the vector \eqref{eq:full-diffeo} by \eqref{eq:xi-r0}, \eqref{eq:F-def}, and \eqref{eq:xi-r1}. Computing how the rest of the components of the metric transform, using the transformation property of the metric under linearized diffeomorphisms 
\be 
\label{eq:shock-mem}
g^{\rm shock}_{\mu\nu} = g^{\rm mem}_{\mu\nu} + \nabla_{(\mu} \xi_{\nu)}  ,
\ee
we finally arrive at the metric
\begin{align}\label{eq:shock-metric}
\begin{split}
	ds^2_\shock &= -2\,du\,dr + 2 r^2 \,dz\,d\bz + 4\p_z^2 C\p_\bz^2 C \, dz\,d\bz - \Big[ 2r \p_z^2 C \,dz^2 + \cc \Big] \\
	&\qquad -  \bigg[ \p_z\p_\bz N \delta(u-u_0) - \frac{2m_B}{r} -\frac{1}{r}\bigg( 2m_B N - \frac{1}{2} \big(\p_z^2 N \p_\bz^2 C + \cc \big)  \bigg) \delta(u-u_0) \bigg] du^2 \\
	&\qquad - \bigg[  N \delta(u-u_0) - \frac{1}{r^2} \p_z^2 C \p_\bz^2 C \bigg] du\,dr - \bigg[ \Big( r\p_z N \delta(u-u_0) + 2\p_z^2\p_\bz C \\
	&\qquad - \big( \p_\bz N \p_z^2 C - N\p_z^2\p_\bz C \big) \delta(u-u_0) \Big) du\,dz + \cc \bigg]  . 
\end{split}
\end{align}
We will also derive a generalization of this metric directly by imposing the vacuum Einstein equation in Appendix~\ref{app:shockwave-met}.

Thus, we see that there is a delta function in $u$ parametrizing the $\mathcal{O}(1)$ component of $g_{uu}$, namely $\partial_z \partial_{\bar z} N \delta(u-u_0)$, which is characteristic of shockwave geometries.  However, note that there are additional metric components $g_{ur},g_{uz}$ that also do not fall off in the large-$r$ limit, and are not usually considered in shockwave metrics.  We show in the next two subsections that these additional components together with $m_B \neq 0$ allow for the vacuum Einstein equation to be satisfied (unlike the standard shockwave metric which is sourced by matter). Furthermore, these terms do not change the symplectic structure of the spacetime, and hence the resulting commutators, in comparison to the memory metric.

\subsection{'t Hooft commutators for $m_B = 0$}\label{ssec:mB0}

In this subsection, we review the planar null coordinate version of the special case studied in \cite{He:2023qha} where the $\CO(r^{-1})$ term in $g_{uu}$ vanishes. Setting $m_B = 0$ and working only to linear order in the soft variables, the shockwave metric \eqref{eq:shock-metric} becomes
\begin{align}\label{AS-met-modified}
\begin{split}
	ds^2_\shock &= -2\,du\,dr + 2 r^2 \,dz\,d\bz  - \p_z\p_\bz N \delta(u-u_0) \, du^2 - N \delta(u-u_0) \, du\,dr \\
	&\qquad - \bigg[ \Big( r\p_z N \delta(u-u_0) + 2\p_z^2\p_\bz C  \Big) du\,dz + \cc \bigg]  - \Big[ 2r \p_z^2 C \,dz^2 + \cc \Big] + \cdots .
\end{split}	
\end{align}
This is precisely the shockwave metric studied in \cite{He:2023qha}, with the shockwave parameter $\alpha$, or the $\CO(r^0)$ component of $g_{uu}$, taking the form\footnote{In \cite{He:2023qha}, $C_{zz}^\text{vac} = -2 \p_z^2 C$ in planar null coordinates.}
\begin{align}\label{eq:alpha}
\begin{split}
	\alpha(z,\bz) = - \p_z\p_\bz N(z,\bz)  = - \frac{1}{2} \Box N(z,\bz).
\end{split}
\end{align}
We see that the memory mode $N$ is identified with the shockwave parameter.

Reviewing the analysis done in \cite{He:2023qha} in our new coordinates, we can solve the $uu$-component of the Einstein equation in the presence of matter to leading order in a large-$r$ expansion to be
\be 
\label{eq:EE-matter}
	\frac{1}{4} \Box^2 N \delta(u - u_0)  + \mathcal{O}( \delta') =  8\pi G T^{M(2)}_{uu}, \qquad T_{\mu\nu}^{M} \equiv \sum_{k=2}^\infty \frac{T_{\mu\nu}^{M(k)}}{r^k} .
\ee
The $\cdots$ in \eqref{AS-met-modified} involve subleading contributions in a large-$r$ expansion which ensure that the Einstein equation is obeyed order by order in $r^{-1}$. We note the appearance of overleading terms in $g_{ur}$ and $g_{uz}$ respectively at $\mathcal{O}(1)$ and $\mathcal{O}(r)$. These terms are generated by applying the linearized diffeomorphism \eqref{eq:full-diffeo} to \eqref{eq:Mink-vacua} with $m_B = 0$, a soft shear profile, and non-vanishing matter $T_{uu}^{M}$. It is thus guaranteed that \eqref{AS-met-modified} solves the Einstein equation in the presence of matter. We will see in Section \ref{ssec:gen-mB} that \eqref{AS-met-modified} may be completed into a vacuum solution by turning on $m_B$. We also note that \eqref{AS-met-modified} is the spherical counterpart of the exact, planar shockwave solutions studied in \cite{tHooft:1987vrq}. It should be possible to show that \eqref{AS-met-modified} can be obtained from the planar shockwave solutions by the appropriate change of coordinates and an expansion at large-$r$.  

It immediately follows from \eqref{eq:EE-matter} that $N$ is related to one of the shockwave variables \cite{tHooft:1987vrq,tHooft:1996rdg}, namely
\be 
\label{eq:'tHooft}
	P_u(z,\bz) \equiv \int_{-\infty}^u du' \, T_{uu}^{M(2)}(u',z, \bz) = \frac{1}{32\pi G_N} \Box^2 N ,
\ee
for $u > u_0$. This identification together with \eqref{eq:soft-comm} led \cite{He:2023qha} to postulate that the 't Hooft commutator from \cite{tHooft:1996rdg} is precisely the soft commutator \eqref{eq:soft-comm} on the gravitational phase space of AFSs, where the variable $X^u$ conjugate to $P_u$ is identified with the Goldstone mode $C(z,\bz)$ via
\be
\label{eq:XC}
	X^u(z,\bz) = -C(z,\bz).
\ee
Direct substitution of \eqref{eq:'tHooft} and \eqref{eq:XC} into the soft commutation relation \eqref{eq:soft-comm} leads to the 't Hooft commutator
\be 
\label{eq:tHooft-comm}
	\big[P_u(z,\bz), X^u(w,\bw) \big] = -i \delta^{(2)}(z - w).
\ee
In the next subsection, we will generalize this analysis to the case when $m_B \neq 0$.

\subsection{Shockwave commutators for $m_B \neq 0$}\label{ssec:gen-mB}

We now propose a generalization of the shockwave commutation relations for the shockwave solution \eqref{eq:shock-metric} with $m_B \neq 0$.  In contrast to \eqref{AS-met-modified}, which can be seen from \eqref{eq:EE-matter} to satisfy the $uu$-component of the Einstein equation with non-vanishing $N$ only in the presence of matter, \eqref{eq:shock-metric} can obey the Einstein equation even in the absence of matter for an appropriate choice of $m_B$. As such, we will see that given a soft/memory profile for the shear, the Einstein equation can be obeyed either by turning on matter sources, or by turning on the Bondi mass aspect. Similar observations apply to Maxwell theory in Minkowski space, and have previously been discussed in \cite{Kapec:2017tkm, Chen:2023tvj, Chen:2024kuq}. We now proceed to show that on-shell, the field-dependent diffeomorphisms \eqref{eq:full-diffeo} with the special $\xi$ given in \eqref{eq:F-def} preserve the symplectic form.

Let us begin by studying the equations of motion for the shockwave metric given a nonvanishing $m_B$. We can rewrite the shockwave metric \eqref{eq:shock-metric} as
\begin{align}\label{eq:shock-met2}
\begin{split}
	ds^2_\shock &= -2\,du\,dr + 2 r^2 \,dz\,d\bz  + 4\p_z^2 C\p_\bz^2 C \, dz\,d\bz - \Big[ 2r \p_z^2 C \,dz^2 + \cc \Big] \\
	&\qquad  \bigg[ - \p_z\p_\bz N \delta(u-u_0) + \frac{2m_B'}{r} \bigg] du^2 - \bigg[ N \delta(u-u_0) - \frac{1}{r^2} \p_z^2 C \p_\bz^2 C \bigg] du\,dr \\
	&\qquad - \bigg[ \Big( r\p_z N \delta(u-u_0) + 2\p_z^2\p_\bz C  - \big( \p_\bz N \p_z^2 C - N\p_z^2\p_\bz C \big) \delta(u-u_0) \Big) du\,dz + \cc \bigg]  ,
\end{split}	
\end{align}  
where $m_B'$ is the shockwave Bondi mass aspect and is related to the memory Bondi aspect to linear order in $N$ by the Lie derivative
\begin{align}\label{eq:shock-mB}
\begin{split}
	m_B' \equiv m_B + \mathcal{L}_{\xi} m_B =  m_B + \bigg( m_B N  - \frac{1}{4} \big( \p_z^2 N \p_\bz^2 C + \cc \big) \bigg) \delta(u-u_0). 
\end{split}
\end{align}
Substituting \eqref{eq:shock-met2} into the $uu$-component of the Einstein equation in the presence of matter, we get to linear order in $N$
\begin{align}\label{eq:mB-shock-eqn0}
\begin{split}
	&-2\p_u m_B' + \big( \p_z^2\p_\bz^2 N + N \p_u m_B'  \big) \delta(u-u_0) \\
	&\qquad\qquad +  \left( 2m_B' N - \frac{1}{2} \p_z^2 C \p_{\bz}^2 N + \cc \right) \delta'(u - u_0) = 8\pi G_N T_{uu}^{M(2)}.
\end{split}
\end{align}
Upon substituting in $m_B'$ given \eqref{eq:shock-mB}, we find
\begin{align}\label{eq:mB-shock-b}
	-2\p_u m_B + \big( \p_z^2 \p_\bz^2 N - \p_u m_B N ) \delta(u-u_0) = 8 \pi G_N T_{uu}^{M(2)}.
\end{align}
This automatically implies in the absence of matter that,
\begin{align}\label{eq:EE-shock}
	\p_u m_B = \frac{1}{2} \p_z^2 \p_\bz^2 N \delta(u-u_0) + \CO(N^2) .
\end{align}
Note that this is precisely \eqref{eq:afs-constraint} to linear order in $N$, which is of course to be expected since linearized diffeomorphisms leave the equations of motion invariant.

To determine what the shockwave commutator should be in the presence of a nontrivial $m_B$ (and hence nontrivial $m_B'$ by \eqref{eq:shock-mB}), we will first show that the symplectic form remains unchanged under the diffeomorphism relating the memory and shockwave metrics. From this, we can deduce that the soft commutator \eqref{eq:soft-comm} still holds. Assuming that we can still identify $X^u$ with the Goldstone $C$ given in \eqref{eq:XC}, we obtain a relation between $N$ and the shockwave momentum $P_u$  generalizing the 't Hooft commutator to the case of non-vanishing Bondi mass.

To determine the symplectic form for the shockwave metric, we substitute \eqref{eq:shock-met2} into \eqref{eq:symp-pot}. To linear order in $N$ (yet keeping terms bilinear in $N$ and $C$), we obtain
\begin{align}\label{eq:shock-theta}
\begin{split}
	\th^u_\shock &= -\frac{1}{r} \delta N \delta(u-u_0) + \CO(r^{-3}) \\
	\th^r_\shock &= \frac{1}{2}\delta N \delta'(u-u_0) - \frac{1}{r}\p_z\p_\bz \delta N \delta(u-u_0) + \frac{1}{r^2} \bigg[ 2 \delta m_B' + 2 \p_z^2 \p_\bz^2 \delta C \\
	&\qquad - \big( 2 N \delta m_B' + 3 m_B' \delta N \big) \delta(u-u_0)   - \frac{1}{2} \delta(u-u_0) \big( \p_z \delta N \p_z \p_\bz^2 C - \p_\bz^2 \delta N \p_z^2 C \\
	&\qquad + \p_z^2 N \p_\bz^2 \delta C + \cc \big)   \bigg] + \CO(r^{-3}) \\
	\th^z_\shock &= \frac{1}{2r^2} \p_\bz \delta N \delta(u-u_0) + \CO(r^{-3}) .
\end{split}
\end{align}
As in the memory metric case, only $\th^r$ contributes when we integrate over $\ci^+$, so the symplectic potential on $\ci^+$ is given by
\be \label{eq:shock-symp}
\begin{split}
	\Th^+_\shock(g,\delta g) &= \frac{1}{16\pi G_N} \int_{\ci^+} du\,d^2z \, \sqrt{-g} \th^r_\shock(g, \delta g) \\
	&=  \frac{1}{16\pi G_N}\int_{\ci^+} du\,d^2z \bigg[  2 \delta m_B'  + \delta(u-u_0)  \Big[ 2 \p_z^2 C  \p_\bz^2 \delta N - \delta( N  m_B') - 2 m_B' \delta N  \Big]  \bigg] ,  
\end{split}
\ee
where we used the fact the determinant factor for the shockwave is
\begin{align}\label{eq:shock-det}
	\sqrt{-g} = r^2 + \frac{r^2}{2} N \delta(u-u_0) - \frac{1}{2} \p_z^2 C \p_\bz^2 C + \CO(r^{-1}),
\end{align}
as well as the fact the $u$ integral eliminates any terms proportional to $\delta'(u-u_0)$, and the planar integral eliminates any total derivatives in $z,\bz$.

We can now use \eqref{eq:symp-form-def} to compute the symplectic form for the shockwave metric. By using the constraint equation \eqref{eq:EE-shock}, and assuming that $m_B$ is independent on $C$,\footnote{In other words, we assume that $C$ will not appear as an integration constant when solving for $m_B$ in \eqref{eq:EE-shock}. This is always the case in theories of where the gravitational contribution to $m_B$ vanishes at the future boundary $\ci^{+}_+$ of $\ci^+$. \label{fn:C-indep}} it follows from \eqref{eq:shock-mB} that all terms involving $m_B'$ will not contribute to the symplectic form, as they contribute either a total phase space variation or terms cubic in the fields. Hence, we find the symplectic form\footnote{Since the result does not depend on the cut of $\mathscr{I}^+$, we can choose this to be $\mathscr{I}^+_-$.} 
\be 
\begin{split}
\label{eq:shock-symp-form}
	 \Omega_\shock^+ &=  \frac{1}{8\pi G_N} \int_{\ci^+_-} d^2z \,  \delta \p_z^2 C \wedge  \delta \p_\bz^2  N  = - \frac{1}{8\pi G_N} \int_{\ci^+_-} d^2z \, \delta \p_\bz^2  N \wedge \delta \p_z^2 C ,    
\end{split}
\ee
where we used integration by parts to simplify the expression, as well as the antisymmetry of the wedge product. This precisely agrees with the memory symplectic form on the soft sector of the gravitational phase space \eqref{eq:mem-sf}.

One way to understand this result is as follows. Denoting $g_{\mu\nu}^{(k)}$ to be the $\CO(r^{-k})$ component of $g_{\mu\nu}$, the phase space of the larger family of spacetimes, with both the shockwave parameter $\alpha \equiv g_{uu}^{(0)} = \Box g_{ur}^{(0)} + \mathcal{O}(N^2)$ and the memory shear \eqref{eq:soft-shear} turned on, is parameterized by $C$ and $-2 \p_z^2 g_{ur}^{(0)} + \p_u g_{zz}^{(-1)}$. Indeed, the combination $-2\p_z^2 g_{ur}^{(0)} + \p_u g_{zz}^{(-1)}$ is invariant under the diffeomorphisms \eqref{eq:full-diffeo}, and the memory and shockwave metrics are respectively obtained by using these diffeomorphisms to fix either $g_{ur}^{(0)} = -1$ or $g_{zz}^{(-1)} = -2\p_z^2 C$. 

As the symplectic form \eqref{eq:shock-symp-form} remains unchanged, it follows that in order to recover the 't Hooft commutator in the presence of a nontrivial Bondi mass, we again require the shockwave momentum to be given in terms of the shockwave parameter $\a(z,\bz)$ in \eqref{eq:alpha}, namely 
\be \label{eq:Prequire}
	P_u(z, \bz) =  \frac{1}{32\pi G_N} \Box^2 N .
\ee
The diffeomorphism relating \eqref{eq:shock-metric} to \eqref{eq:Mink-vacua} hence shows that $P_u$ is to be identified with the soft charge aspect $q_s$, namely \cite{Strominger:2017zoo}
\be 
\label{eq:soft-ch}
	q_s \equiv -\frac{1}{8}\Box^2 N \implies P_u = -\frac{ q_s}{4\pi G_N}.
\ee 
However, notice that this implies the first equality in \eqref{eq:'tHooft} no longer holds, since by \eqref{eq:mB-shock-b} $T_{uu}^{M(2)}$ includes contributions from $m_B$. Indeed, we see that \eqref{eq:mB-shock-b} and \eqref{eq:Prequire} imply to linear order in $N$ 
\begin{align}\label{eq:Pminus-bondi}
\begin{split}
	P_u(z,\bz) &=  \int_{-\infty}^u du\, T_{uu}^{M(2)} + \frac{1}{4\pi G_N} \Delta m_B,
\end{split}
\end{align}
for $u > u_0$ and $\Delta m_B \equiv m_B(u) - m_B(-\infty)$. Particularly interesting is the case of the vacuum, where $T_{uu}^{M} = 0$ and $m_B \not= 0$, in which case \eqref{eq:Pminus-bondi} becomes
\begin{align}
\begin{split}
	P_u(z,\bz) = \frac{1}{4\pi G_N} \Delta m_B.
\end{split}
\end{align}
We hence see that the Bondi mass here plays a similar role as the integrated matter stress tensor in the case of the shockwave metric \eqref{AS-met-modified} with $m_B = 0$. In other words, the constraint \eqref{eq:EE-shock} shows that $m_B$ sources the memory mode even in absence of matter. We will comment further on the relation between spacetime fluctuations and fluctuations in the Bondi mass in Section \ref{sec:modular-Hamiltonian}.

\section{Boundary actions in gravity}

\label{sec:bdry-act}

In this section, we derive an on-shell action for the infrared sector of gravitational theories in AFSs. We show that this action localizes to a codimension-2 surface -- the ``celestial'' space \cite{Strominger:2017zoo} or the corner of a causal diamond \cite{Freidel:2020xyx,Freidel:2020svx,Freidel:2020ayo,Freidel:2021cjp}. We then study the transformation of this action under linearized diffeomorphisms generated by the vector field \eqref{eq:full-diffeo}. This transformation can be derived abstractly from the transformation of the symplectic potential under diffemorphisms \cite{Donnelly:2016auv}, or by explicit comparison with the on-shell action associated with the shockwave metric \eqref{eq:shock-metric}. Utilizing the formalism developed in \cite{Donnelly:2016auv}, we find that the memory on-shell action can be made gauge-invariant by the addition of a corner term. The resulting on-shell action hence characterizes both AFSs and shockwave geometries related by a diffeomorphism.

The on-shell action is obtained by demanding that the variational principle holds. Thus, we begin by varying the EH action without any boundary terms and imposing the vacuum Einstein equation. Fortunately, as is illustrated in Appendix~\ref{app:symp-pot}, this is precisely the definition of the symplectic potential, given in  \eqref{eq:symp-pot} to be 
\begin{align}\label{eq:EH-var}
	\delta S_{\EH} = \Th(g,\delta g) = \frac{1}{16\pi G_N} \int_{\p \CM} d\Sigma_\s \, \th^\s(g,\delta g).
\end{align}
It is immediate to see from \eqref{eq:AF-sp} that in the presence of gravitational flux, so that $N_{AB} \neq 0$, the variation is not only non-vanishing, but also not an exact form on phase space \cite{Wald:1999wa,Fiorucci:2021pha}. 
For the special case given in \eqref{eq:soft-shear}, where the shear only consists of soft modes, we will see that this can be remedied by an appropriate choice of boundary conditions, which will render the variational principle well-defined,\footnote{The reason for this is that in the soft sector, the Goldstone and the memory modes are independent. In particular,  fixing the zero mode of the shear doesn't imply that the news (or leading memory mode) vanishes. Nevertheless in this case one can see from \eqref{eq:delta-mem1} that the variational principle becomes well defined.} and the addition of a boundary term to the action.
Since $S_{\rm EH} = 0$ on-shell, the boundary action will then coincide with the on-shell action.

In Sections~\ref{ssec:mem-OS} and \ref{ssec:shock-OS}, we will respectively obtain the on-shell action for memory and shockwave spacetimes. Next, in Section~\ref{ssec:GI-OS}, we will determine an additional boundary term that we can add to render the on-shell action invariant under the diffeomorphisms generated by \eqref{eq:full-diffeo}. Finally, note that thus far, all our analysis has been focused on $\ci^+$. To account for the full boundary of the spacetime, we need to also include the contribution from $\ci^-$. This is remedied in Section~\ref{ssec:matching}, where we use the standard matching condition from \cite{Strominger:2013jfa} to obtain the $\ci^-$ contribution to the on-shell action.

\subsection{On-shell action in memory spacetimes}\label{ssec:mem-OS}

We begin by computing \eqref{eq:EH-var} for the memory metric. Using \eqref{eq:symp-pot-coords}, we get on $\ci^+$
\begin{align}\label{eq:delta-mem1}
\begin{split}
	\delta S^{\mem +}_{\EH} &= \frac{1}{32\pi G_N} \int_{\ci^+} du\,d^2z \bigg[ 4 \delta m_B  + \frac{1}{4} \delta \big( \p_z^2 N \p_\bz^2 N  \big) \delta(u-u_0)  \\
	&\qquad + \frac{1}{2} \delta \big(  \p_z^2 C \p_\bz^2 N  + \cc \big)\delta(u-u_0) - 2\big( \p_z^2 \delta C \p_\bz^2 N + \cc \big) \delta(u-u_0)  \bigg] ,
\end{split}
\end{align}
where the superscript $+$ indicates we are evaluating the boundary action contribution from $\ci^+$; there is an analogous contribution from $\ci^-$ that we will include later in Section~\ref{ssec:matching}. We now impose the boundary conditions\footnote{The memory mode $N$ measures the change in $C_{zz}$ between $\ci^+_-$ and $\ci^+_+$ and may in general vary. This is because in any scattering process, it is the net supertranslation charge, which consists of both memory and hard components, that is typically assumed to be conserved across spatial infinity (see \cite{Strominger:2017zoo} for a review). Fixing $N$ would therefore correspond to fixing the flux or hard charge as well, which seems overly restrictive. Note also that our boundary condition fixes the supertranslation frame at spatial infinity, and such a choice is implicit in all computations of gravitational waveforms \cite{Mitman:2024uss}.} 
\be
\label{eq:deltaC}
\delta \p_z^2 C = 0, \quad \delta\p_{\bz}^2 C = 0 .
\ee 
This boundary condition \ref{eq:deltaC} is preserved by diffeomorphisms parameterized by \eqref{eq:full-diffeo} with $F$ given in \eqref{eq:F-def}. This will be important later, since the on-shell action for this boundary condition inherits the transformation properties of the symplectic potential. 

Imposing this boundary condition, \eqref{eq:delta-mem1} becomes a total variation that can be cancelled by the addition of the boundary action 
\be 
\begin{split}
	 S_\bdy^{\mem +} &= - \frac{1}{32\pi G_N} \int_{\ci^+} du\,d^2z \bigg[ 4 m_B   + \frac{1}{2}  \big(  \p_z^2 C \p_\bz^2 N  + \cc \big)\delta(u-u_0) + \mathcal{O}(N^2)  \bigg]. 
\end{split}
\ee
Recall that we are working to linear order in $N$. Now, observe from \eqref{eq:mB-constraint} that $\p_u m_B$ is to linear order in $N$ a total derivative in $z,\bz$.\footnote{In general, the $m_B$ contribution can be traded via the constraint \eqref{eq:mB-constraint} for an $\mathcal{O}(N^2)$ term with a divergent coefficient. We expect this contribution to be related to infrared divergences, which we discuss further in Section \ref{sec:discussion}.} If we also assume that $m_B$ vanishes at $\mathscr{I}^+_+$, then the integration constant that arises from integrating \eqref{eq:mB-constraint} over $u$ is also a total derivative in $z,\bz$.  This means $m_B$ is a total derivative in $z,\bz$ and hence vanishes when integrated over the transverse space. It follows that
\begin{align}\label{eq:mem-bdy-action}
\begin{split}
	S^{\mem+}_\bdy &= - \frac{1}{64\pi G_N} \int_{\ci^+_-} d^2z \big(  \p_z^2 C \p_\bz^2 N  + \cc \big)  = - \frac{1}{32\pi G_N} \int_{\ci^+_-} d^2z\,    \p_z^2 C \p_\bz^2 N  .
\end{split}
\end{align}

We can compare and contrast this result with one of the main entries of the AdS/CFT dictionary, namely that the on-shell gravity action acts as the generating function for the dual CFT correlators, where the boundary values of the non-normalizable modes $\phi_0$ act as sources \cite{Aharony:1999ti}:
\be 
	Z_{\text{on-shell}}[\phi_0] = \left\langle e^{\int \phi_0 \mathcal{O}} \right\rangle.
\ee
Similarly, \eqref{eq:mem-bdy-action} suggests that in four-dimensional AFSs, the on-shell action truncated to the leading soft/memory sector with $\delta \p_z^2 C = \delta \p_{\bz}^2 C = 0$ generates correlation functions of effectively soft currents $J_{zz} \propto \p_{z}^2 N$ in a two-dimensional (celestial) CFT:
\be 
	Z_{\text{on-shell}}[C] = \left\langle e^{ S_{\rm bdy}^{\rm GI}[C, N]} \right\rangle,
\ee
where $S_{\rm bdy}^{\rm GI}$ was given in \eqref{eq:gi-bdry-intro} and will be derived below. We expect this on-shell action to be corrected by a whole tower of subleading soft contributions \cite{Freidel:2022skz}. In principle, a complete understanding of these corrections would allow us to formulate a dictionary for holographic observables in AFSs relating the bulk on-shell path integral to celestial correlators with sources for an infinite tower of currents turned on. We leave an exploration of these ideas to future work.

\subsection{On-shell action in shockwave spacetimes}\label{ssec:shock-OS}
 
As in the previous subsection, we begin by computing \eqref{eq:EH-var} for the shockwave metric. Using \eqref{eq:shock-symp}, we get on $\ci^+$
\begin{align}\label{delta-shock0}
\begin{split}
	\delta S_\EH^{\shock+} &= \frac{1}{16\pi G_N}\int_{\ci^+} du\,d^2z \bigg[  2 \delta m_B'  + \delta(u-u_0)  \Big[ 2 \p_z^2 C  \p_\bz^2 \delta N - \delta( N \delta m_B') - 2 m_B' \delta N  \Big]  \bigg] \\
	&= \frac{1}{8\pi G_N}\int_{\ci^+} du\,d^2z \Big[   \delta m_B'  +   \p_z^2 C \p_\bz^2  \delta N \delta(u-u_0)   \Big] \\
	&= \frac{1}{8\pi G_N}\int_{\ci^+} du\,d^2z \Big[   \delta \big( m_B'  + \p_z^2 C \p_\bz^2 N \delta(u-u_0) \big) -   \p_z^2 \delta C \p_\bz^2  N \delta(u-u_0)   \Big] ,
\end{split}
\end{align}
where in the second line we recalled from Footnote~\ref{fn:C-indep} that $m_B$ is independent of $C$, which implies by \eqref{eq:shock-mB} that $m_B'$ is only linear in $N$, and so any terms involving a product of $N$ and $m_B'$ in the above equation can be dropped.
In the last equality we rewrote the integrand so the term that is not a total variation involves $\delta C$ rather than $\delta N$. 

Imposing the boundary condition \eqref{eq:deltaC}, we are once again left with a total variation term. The variational principle holds if this total variation can be canceled, which is achieved if we add to the EH action
\begin{align}\label{eq:bdry-shock}
\begin{split}
	S_\bdy^{\shock+} &= - \frac{1}{8\pi G_N}\int_{\ci^+} du\,d^2z   \big( m_B'  + \p_z^2 C \p_\bz^2 N \delta(u-u_0) \big)    \\
	&= - \frac{1}{8\pi G_N}\int_{\ci^+} du\,d^2z   \bigg[ m_B + \bigg( m_B N - \frac{1}{4} \big( \p_z^2 N \p_\bz^2 C + \cc \big) \bigg) \delta(u-u_0) \\
	&\qquad  + \p_z^2 C \p_\bz^2 N \delta(u-u_0) \bigg]     \\
	&= - \frac{1}{16\pi G_N}\int_{\ci^+_-} d^2z \,  \p_z^2 C \p_\bz^2 N   ,
\end{split}
\end{align}
where in the second equality we substituted in \eqref{eq:shock-mB}, and in the last equality we dropped all $\CO(N^2)$ terms (recall $m_B = \CO(N)$) and then integrated over $u$. Comparing with the memory boundary action \eqref{eq:mem-bdy-action}, we see that they differ by a factor of $2$. This is expected since the symplectic potentials, and hence the on-shell actions, are in general not diffeomorphism-invariant. This has already been observed in \cite{Donnelly:2016auv} and was remedied by adding a corner term to the EH action. We will next describe this procedure for the special case of the diffeomorphism relating the shockwave and memory metrics.

\subsection{Diffeomorphisms and an invariant on-shell action}\label{ssec:GI-OS}

A diffeomorphism is a smooth, invertible map between two manifolds. We will be interested in the case where $\phi$ maps the spacetime manifold $\CM$ to itself, i.e.,
\be 
	\phi: \CM \rightarrow \CM.
\ee
The map $\phi$ induces the pullback map $\phi^*$, which acts on forms $\Omega \in T^* \CM$ on the cotangent space, and the pushforward map $\phi_*$, which acts on vectors $V \in T\CM$ on the tangent space. Henceforth, we will be using lowercase Latin letters to denote abstract tensor indices.

Now, the symplectic potential density  $(\star\theta)_{abc}$, where $\theta_a$ is the spacetime 1-form whose components appear in the integrand of \eqref{eq:symp-pot} and $\star$ is the Hodge star, is a 1-form on phase space and a $3$-form in four-dimensional spacetime. The transformation of the symplectic potential density $\star\theta$ takes the form \cite{Donnelly:2016auv}
\be 
\label{eq:diff-transform-pot}
	\star \theta(\phi^*g, \delta(\phi^*g)) = \phi^*\left( \star\theta(g, \delta g) + \star\theta(g, \mathcal{L}_{\delta_{\phi}} g)\right),
\ee
where $\mathcal{L}$ is the spacetime Lie derivative, and $\delta_{\phi}$ is a spacetime vector (but a phase space one-form) given by
\be 
\delta_{\phi} = \delta \phi^a \circ \phi^{-1} \p_a.
\ee
As shown in \cite{Donnelly:2016auv}, the second term in \eqref{eq:diff-transform-pot} is due to the non-trivial transformation properties of $\phi$ under field space variations and is only present for field-dependent diffeomorphisms. This is precisely the case for us, as the function $F$ parametrizing the diffeomorphism vector $\xi^a$ \eqref{eq:full-diffeo} is given by \eqref{eq:F-def}, which depends on $N$ and hence takes values in the soft phase space.  As an explicit check, we show in Appendix~\ref{app:diff-action-symp-pot} that the difference between the memory and shockwave symplectic potentials, given respectively in \eqref{eq:soft-symp} and \eqref{eq:shock-symp}, is indeed consistent with \eqref{eq:diff-transform-pot} for the special case of a linearized diffeomorphism
\be 
	\phi^a(x) = x^a + \xi^a,
\ee
where $\xi^a$ is given in \eqref{eq:full-diffeo}. 

Of course, the transformation property \eqref{eq:diff-transform-pot} of the symplectic potential under diffeomorphisms implies the on-shell action \eqref{eq:mem-bdy-action} is also not invariant under diffeomorphisms. However, we now demonstrate that by adding the additional boundary term, given by
\begin{align}\label{eq:Scorner}
	S_{\p}^+ &=  -\frac{1}{16\pi G_N} \int_{\ci^+} du\,d^2z\, \frac{1}{8} \big( C_{zz} N_{\bz\bz} + \cc \big) ,
\end{align}
to the EH action, we can obtain an on-shell action that is invariant under the diffeomorphism considered in this paper. The role of this term is to transform under the linearized diffeomorphism relating the memory and shockwave metrics precisely in such a way to cancel the associated change in the boundary action induced by the second term in \eqref{eq:diff-transform-pot}. The addition of (the negative of) \eqref{eq:Scorner} to \eqref{eq:mem-bdy-action} should therefore lead to a gauge-invariant on-shell action. 

Notice that the extra boundary action is entirely determined by the shear and news tensor of the metric. For the case of the memory metric, we can substitute in the memory shear profile \eqref{eq:soft-shear} to get 
\begin{align}
\begin{split}
	S_\p^{\mem+} &= \frac{1}{32\pi G_N} \int_{\ci^+_-} d^2z\,    \p_z^2 C \p_\bz^2 N  + \CO(N^2) ,
\end{split}
\end{align}
where we combined terms by integrating by parts. Notice that this is a \emph{corner action} because it is completely localized on the celestial sphere at $\ci^+_-$. With the inclusion of this extra corner action, the on-shell action for the memory metric \eqref{eq:mem-bdy-action} is suitably modified to
\begin{align}\label{eq:mem-os-new}
\begin{split}
	S_{\bdy}^{\mem+} - S^{\mem+}_\p = - \frac{1}{16\pi G_N } \int_{\ci^+_-} d^2z\, \p_z^2 C \p_\bz^2 N .
\end{split}
\end{align}

Turning now to the shockwave on-shell action, recall that we obtain the shockwave metric \eqref{eq:shock-met2} by eliminating the memory mode (to linear order in $N$) using the diffeomorphism generated by $\xi^a$ given in \eqref{eq:full-diffeo}. As $\xi^a$ was constructed to remove the memory mode so that $C_{zz} \to -2\p_z^2 C$, this implies $N_{zz} = 0$. Substituting this into \eqref{eq:Scorner} yields
\begin{align}
	S_\p^{\shock+} = 0,
\end{align}
and so the on-shell action for the shockwave metric \eqref{eq:bdry-shock} is not modified. Indeed, it is precisely equal to \eqref{eq:mem-os-new}, indicating that the modified on-shell action is invariant under the diffeomorphism that takes the memory metric to the shockwave metric. We will therefore label the gauge-invariant on-shell action as 
\begin{align}\label{eq:S-gi}
	S^{\GI+}_\bdy &\equiv - \frac{1}{16\pi G_N} \int_{\ci^+_-} d^2z\,\p_z^2 C \p_\bz^2 N .
\end{align}
Note that in this formula we have combined the on-shell bulk action \eqref{eq:bdry-shock} with the corner term \eqref{eq:Scorner}. The transformation of \eqref{eq:S-gi} under diffeomorphisms is now obscured and is not simply given by a shift in $N$.  Our addition of the corner term realizes the general construction of \cite{Donnelly:2016auv} for the case of the simplest asymptotically flat background that carries gravitational memory, restricted to only terms linear in $N$. We will comment on the structure of the terms quadratic in $N$ in Section \ref{sec:discussion}.

\subsection{Matching condition}\label{ssec:matching}

Thus far, we have only described the contribution to the on-shell action coming from $\ci^+$. Indeed, as we see in \eqref{eq:EH-var}, the variation of the action receives contributions from the boundary of the spacetime manifold, which includes both $\ci^+$ and $\ci^-$. In this subsection, we will account for the contribution from $\ci^-$. Fortunately, the analysis almost exactly parallels the $\ci^+$ case, and hence we will not need to repeat it. The analog of the gauge-invariant on-shell action \eqref{eq:S-gi} on $\ci^-$ is to linear order in $N$ given by
\be \label{eq:S-gi-minus}
S_\bdy^{\GI -}   = \frac{1}{16\pi G_N}  \int_{\ci^-_+} d^2z \, \p_z^2 C \p_{\bz}^2 N + \mathcal{O}(N^2) ,
\ee
where we have a relative minus sign from \eqref{eq:S-gi} due to the fact $\ci^+$ and $\ci^-$ have opposite orientations. We now use the standard matching conditions \cite{He:2014laa}
\begin{align}
\begin{split}
	C_{zz}\big|_{\ci^+_-} = C_{zz}\big|_{\ci^-_+},
\end{split}
\end{align}
which upon substituting in the memory shear profile \eqref{eq:soft-shear} yields the boundary condition
\be 
\begin{split}\label{eq:matching}
	C \big|_{\ci^+_-} &= C \big|_{\ci^-_+} .
\end{split}
\ee
This amounts to imposing a no-flux through $i^0$ boundary condition. 
Furthermore, 
\be 
\begin{split}\label{eq:matching}
	 N \big|_{\ci^+_-}  = - N \big|_{\ci^-_+} ,
\end{split}
\ee
which is a consequence of the properties of soft gravitons  under crossing \cite{He:2014laa, Strominger:2017zoo}. We conclude that the full gauge-invariant on-shell action to leading order in $N$ is 
\begin{align}\label{eq:S-os-total}
\begin{split}
	S_\bdy^\GI &\equiv S^{\GI+}_\bdy + S^{\GI-}_\bdy \\
	&= - \frac{1}{16\pi G_N} \int_{\ci^+_-} d^2z\,\p_z^2 C \p_\bz^2 N + \frac{1}{16\pi G_N} \int_{\ci^-_+} d^2z\, \p_z^2 C \p_\bz^2 N \\
	&= - \frac{1}{8\pi G_N} \int_{\ci^+_-} d^2z\,\p_z^2 C \p_\bz^2 N ,
\end{split}
\end{align}
where in the second line we substituted in \eqref{eq:S-gi} and \eqref{eq:S-gi-minus}, and in the last line we used the matching condition \eqref{eq:matching}.

\section{Three equivalent interpretations of the on-shell action}
\label{sec:equiv-actions}

The goal of this section is to show that the on-shell EH action in the soft limit derived in the previous section -- the gauge-invariant on-shell action given in \eqref{eq:S-os-total} -- has three different interpretations. In Section~\ref{sec:ST-charge}, we show that the on-shell action is precisely the soft supertranslation charge to linear order in $N$, with the supertranslation parameter identified with the Goldstone mode $C$. Next, in Section~\ref{sec:modular-Hamiltonian}, we demonstrate that the on-shell action can be recast into the Verlinde-Zurek (VZ) shockwave effective action studied in \cite{Verlinde:2022hhs}. Finally, in Section~\ref{sec:shockwave-soft-action}, we further show that the on-shell action is also equivalent to the interaction term in the soft effective action proposed by Kapec and Mitra (KM) in \cite{Kapec:2021eug}, which is to linear order in $N$ the full soft effective action. This provides not only a derivation of the shockwave effective action and the soft effective action, but also establishes a diamond of equivalences in the infrared sector of perturbative gravitational theories among the gauge-invariant on-shell action, the shockwave effective action, the soft effective action, and the soft supertranslation charge to leading order in the memory mode. This work has led to ideas that were explored in the context of Abelian gauge theories in \cite{He:2024ddb, He:2024skc}.

\subsection{On-shell action = soft supertranslation charge}
\label{sec:ST-charge}

We first note that the total gauge-invariant on-shell action is precisely the soft supertranslation charge to linear order in $N$ if we promote the function $f$ parametrizing the supertranslation charge to the Goldstone mode $C$. Explicitly, the supertranslation charge we are considering to quadratic order in fields is in planar null coordinates \cite{He:2014laa} 
\begin{align}\label{eq:ST-charge}
\begin{split}
	Q_{f=C} &= \frac{1}{4\pi G_N} \int_{\ci^+_-} d^2z\, C m_B  = - \frac{1}{16\pi G_N} \int_{\ci^+} du\, d^2z\, C \big( \p_z^2 N^{zz} + \cc \big) ,
\end{split}
\end{align}
where in the last equality we used \eqref{eq:mB-constraint}.  This corresponds to the soft part of the supertranslation charge. If we now specialize to the memory metric \eqref{eq:soft-shear}  and then integrate by parts, we then have 
\begin{align}
\label{eq:soft-charges-C}
\begin{split}
	Q_{f=C}^\mem &= - \frac{1}{8\pi G_N} \int_{\ci^+_-} d^2z\, \p_z^2 C  \p_\bz^2 N  .
\end{split}
\end{align}
Comparing with the gauge-invariant on-shell action \eqref{eq:S-os-total}, we see that they are exactly equal. The $\CO(N)$ contribution to the gauge-invariant on-shell action is hence the soft graviton charge parameterized by a function $C$ that labels the asymptotically flat vacuum near $i^0$!

\subsection{On-shell action = shockwave effective action}
\label{sec:modular-Hamiltonian}

Motivated by \cite{Verlinde:1991iu}, VZ proposed in \cite{Verlinde:2022hhs} an action defined on the bifurcate Killing horizon of a causal diamond in AFSs and connected it to the modular Hamiltonian associated to the diamond. For the four-dimensional case, the action they studied is in our notation given by\footnote{The second equality in \eqref{eq:VZ} was obtained in \cite{Verlinde:2022hhs} by assuming a shockwave geometry where the shockwave is localized near the bifurcate horizon, which in our notation implies $T_{uu}^M \sim \delta(u-u_0)$ with $u_0 = -\infty$. Thus, the identification of the shockwave action with the modular Hamiltonian only holds in this limit.}
\be \label{eq:VZ}
\begin{split}
	 S_\VZ &=  \int_{\CH^+} du\,d^2z\, X^u T_{uu}^{M} =  \int_{S^2} d^2z\, X^u P_u ,
\end{split}
\ee
where $\CH^+$ is the future null horizon (analogous to $\ci^+$), $S^2$ the bifurcate horizon (analogous to $\ci^+_-$), and the second equality follows from \eqref{eq:'tHooft}.\footnote{It was implicitly assumed in \cite{Verlinde:2022hhs} that $m_B$ vanished.} Here $X^u$ and $P_u$ are the shockwave variables defined in terms of the shockwave potential in Section~\ref{sec:thooft}. Recalling the relation between the shockwave variables and soft modes is given by \eqref{eq:XC} and \eqref{eq:Prequire} (even in the presence of nontrivial $m_B$), namely
\be 
	P_u = \frac{1}{32 \pi G_N} \Box^2 N, \qquad X^u = -C,
\ee
we can rewrite \eqref{eq:VZ} as
\be 
\begin{split}
\label{eq:VZ-soft}
	S_\VZ &= - \frac{1}{32\pi G_N} \int_{S^2 } d^2z\, C \Box^2 N = - \frac{1}{8\pi G_N} \int_{S^2} d^2z \,  \p_z^2 C \p_\bz^2 N . 
\end{split}
\ee
Comparing \eqref{eq:S-os-total} with \eqref{eq:VZ-soft}, we see that they precisely coincide in the large-$r$ limit where $S^2 \to \ci^+_-$. 

A few comments are in order. First, we would like to emphasize that in contrast to \cite{Verlinde:2022hhs}, which considers the scattering of two shockwaves in a Minkowski background, the metric \eqref{eq:shock-met2} instead describes \textit{one} shockwave on top of one of the infinity of Minkowski vacua parameterized by $C$. We also recall that in \cite{tHooft:1996rdg} the shifts acquired by each shockwave as it crosses the other were promoted to non-commuting operators, while in \cite{He:2023qha} we considered only one shockwave, but observed that the same commutation relations may be recovered if one identifies the shift of a would-be second shockwave with the background $C$. This resonates with the well-known idea that the leading, eikonal contribution to the scattering amplitude of two high-energy particles can be recovered by considering the propagation of one particle in a background \cite{tHooft:1987vrq, Kabat:1992tb}. That the horizon action \eqref{eq:VZ} expressed in terms of the two shockwave shifts takes the same form as the on-shell action \eqref{eq:S-os-total} for one shock in a $C$ background provides further evidence for a relation along these lines. 

Second, the effective action of \cite{Verlinde:2022hhs} was related to a modular Hamiltonian and is therefore usually associated with hard, rather than soft, degrees of freedom. However, the constraint equation \eqref{eq:mB-shock-b} relates the matter/gravity stress tensor (or hard) degrees of freedom to the soft ones and the Bondi mass aspect (or the supertranslation charge).  The shockwave metrics considered in \cite{Verlinde:2022hhs} have a manifestly vanishing Bondi mass aspect, in which case the hard and soft charge aspects are equal up to a sign. This provides a physical argument for the identification of \eqref{eq:VZ} and \eqref{eq:VZ-soft}. We nevertheless find it intriguing that the same coefficient can be recovered from an \textit{on-shell} action in AFS after imposing a particular  choice of boundary conditions and then adding a corner term that renders the contributions to the on-shell action from $\ci^+$ and $\ci^-$ individually invariant under the diffeomorphism generated by \eqref{eq:full-diffeo}. This derivation is in stark contrast with that of \cite{Verlinde:2022hhs}, which instead constructed an \textit{effective} action by demanding that the equations of motion derived from it agree with the Einstein equation. 

Lastly, we would also like to point out that in the absence of matter, charge conservation implies that $Q_S = Q_S^-$, and hence $S_{\rm bdy}^{\rm GI} = 0$ since a soft graviton insertion vanishes in vacuum in the absence of matter. On the other hand, in the presence of matter, \eqref{eq:VZ-soft} will be non-trivial in special states that carry non-trival soft charges \cite{Freidel:2022skz}, or in soft scattering processes that may \textit{violate} the matching condition (such that $Q_S \neq Q^-_{S}$). We leave the further study of these interesting generalizations to the future.

The rewriting of \eqref{eq:VZ} given in \eqref{eq:VZ-soft} in terms of soft modes now allows us to recast the reasoning of \cite{Verlinde:2022hhs} that led to an estimate of spacetime fluctuations in a quantum theory of gravity in a canonical framework. A key assumption of \cite{Verlinde:2022hhs} is that the boost parameter in \eqref{eq:VZ}, or the Goldstone mode \eqref{eq:VZ-soft}, ought to be promoted to operators in the quantum theory. We do not have a first principle derivation of this assumption at the level of the on-shell action, and indeed allowing for $C$ to be a dynamical variable is naively in conflict with the boundary condition \eqref{eq:deltaC}, which was necessary in our derivation of the on-shell action \eqref{eq:S-os-total}.\footnote{Note also from \eqref{eq:symp-pot-coords} and \eqref{eq:shock-symp} that the variation of the EH action for both the memory and shockwave metrics is \textit{not} phase space exact, and therefore a boundary condition on either $C$, $N$ or a linear combination thereof appears to be necessary in order to satisfy the variational principle.} Nevertheless, we will in the remainder of this subsection postulate that a future theory of quantum gravity will allow us to make sense of this assumption and simply discuss its implications. In this case, the shockwave effective action becomes a bilinear in a two-dimensional quantum theory (or a celestial CFT). We can then consider the normal-ordering prescription
\begin{align}
	\langle 0 | : C\p_z^2\p_\bz^2 N : | 0 \rangle = 0.
\end{align}
The fact the one-point function of $:S_\VZ:$ vanishes suggests the interpretation of this operator as a ``modular fluctuation'' \cite{Verlinde:2022hhs}, so that (see also \cite{Lewkowycz:2013nqa, Liu:2021kay})
\be 
\label{eq:K}
	\Delta K(z,\bz) \equiv \frac{1}{8\pi G_{N}} : C \p_z^2 \p_{\bz}^2 N : \quad\implies\quad  \langle 0| \Delta K |0\rangle = 0.
\ee

Consider now the two-point function of the modular fluctuation $\Delta K$, which will involve the OPE
\be 
\label{eq:bilinear}
\frac{1}{(8\pi G_{N})^2}: C\p_z^2 \p_{\bz}^2 N:(z,\bz) :C \p_{w}^2 \p_{\bw}^2 N: (w,\bw) .
\ee
Recalling the only nontrivial soft commutator is given by \eqref{eq:CN}, we deduce the OPEs
\be \label{eq:OPEs}
\begin{split}
	C(z, \bz) N(w, \bw) &\sim 4 G_N |z - w|^2 \log|z - w|^2, \quad N(z, \bz) N(w, \bw) \sim 0.
\end{split}
\ee
We hence find
\be 
\begin{split}
\Delta K(z,\bz) \Delta K(w, \bw) &\sim \frac{1}{(8\pi G_{N})^2} : C \p_z^2 \p_{\bz}^2  N:(z,\bz)  :C   \p_{w}^2 \p_{\bw}^2 N: (w,\bw)\\
&\sim \delta^{(2)}(z - w) \delta^{(2)}(z - w).
\end{split}
\ee
The two-point function of the global (integrated) modular fluctuation operators is then
\be 
\label{eq:deltaK}
\langle 0| \Delta K^2 |0\rangle \equiv \int d^2z\,d^2w\, \langle 0|\Delta K(z,\bz) \Delta K(w,\bw) |0\rangle = A\delta^{(2)}(0),
\ee
where $A$ is the area of the transverse space, and following \cite{Verlinde:2022hhs}, one may choose to regularize the collinear divergence  $\delta^{(2)}(0)$ by the Planck scale. Noting that the divergent one-point function is computed using \eqref{eq:OPEs}  to be
\be 
\label{eq:K}
\int d^2 z \, \langle 0| K |0\rangle \equiv \frac{1}{8\pi G_N}\int d^2 z \, \langle 0| C \p_z^2 \p_{\bz}^2 N |0\rangle = A \delta^{(2)}(0),
\ee 
the regularization of $\delta^{(2)}(0)$ in \eqref{eq:K} will determine that of \eqref{eq:deltaK}.

A completely analogous computation was outlined in Section 4 of \cite{Verlinde:2022hhs}, where $C$ and $N$ are identified with $X^u$ and $X^v$ (see \cite{He:2023qha}). Thus, the relation between \eqref{eq:VZ} and \eqref{eq:VZ-soft} suggests that the spacetime fluctuations studied in \cite{Verlinde:2022hhs} are directly related to fluctuations in operators obtained from the soft supertranslation charges \eqref{eq:soft-charges-C} by promoting the supertranslation parameter $C$ to an operator. Indeed, we see from \eqref{eq:soft-charges-C} and \eqref{eq:K} that
\be 
	\int_{S^2} d^2z \, K(z,\bz) = Q_{f = C}^{\mem}.
\ee

Before concluding this section, we observe an intriguing relation between the $C, N$ bilinears in \eqref{eq:K} and a spacetime area operator. Such operators can naturally be defined in a celestial CFT \cite{Pasterski:2017kqt, Raclariu:2021zjz, Pasterski:2021rjz, Pasterski:2021raf, Donnay:2023mrd}, and we see evidence here that they may capture holographic information about spacetime area fluctuations. A similar bilinear has appeared in covariant redefinitions of the Bondi mass aspect \cite{Compere:2018ylh, Compere:2020lrt, Donnay:2021wrk}. Related ideas appeared in \cite{Kapec:2016aqd}, including a relation between areas of cuts of null infinity and integrals of the Bondi mass aspect at $\ci^+$ (which may also be related to the superrotation/boost charge via the Einstein equation), while a relation between shockwaves and areas in a covariant phase space formalism has been established in \cite{Liu:2021kay}. Composite operators involving the celestial CFT stress tensor (i.e., subleading counterparts of \eqref{eq:K}) have appeared in \cite{Fiorucci:2023lpb}, as well as \cite{Freidel:2022skz} (at all subleading orders). To the extent of our knowledge, the relation between the $C,N$ bilinear found in \eqref{eq:K} and an area operator is new and it may be worthwhile to understand it and its relation to the vacuum two-point function of the Bondi mass recently computed in \cite{Ciambelli:2024swv} better. Finally, we have focussed on a two-point function computed in the usual vacuum $|0\rangle$ of Minkowski spacetime. Of course, the precise nature of the vacuum state in quantum gravity remains subtle and a computation of a precise observable that probes the vacuum degeneracy in AFSs is beyond the scope of this work. We leave a further exploration of these ideas to the future. 

\subsection{On-shell action = soft effective action}
\label{sec:shockwave-soft-action}

Finally, we point out that the gauge-invariant on-shell action we wrote down in \eqref{eq:S-os-total} agrees at linear order in $N$ with the soft action obtained in \cite{Kapec:2021eug}.  In particular, translating the results of \cite{Kapec:2021eug} to our notation, the interaction term in the soft effective action is in four dimensions given by\footnote{In writing \eqref{eq:soft-eff-init}, we have multiplied the soft effective action found in \cite{Kapec:2021eug} by a factor of $i$. This is because the path integral in \cite{Kapec:2021eug} involves $e^{-S_\KM}$, whereas we are using the convention where the path integral involves $e^{iS_\KM}$ when defining the boundary action.}
\be 
\label{eq:J-soft-eff}
 S_\KM = -\frac{1}{16\pi^2} \int_{\ci^+_-} d^2z\, \CJ_{AB} \widetilde{C}^{AB} + \CO(N^2),
\ee
where $\widetilde{C}^{AB}$ is the shadow transform of $C^{AB}$ in two dimensions, and 
\be 
\label{eq:current}
\CJ_{AB}(z,\bz) = \lim_{\omega \rightarrow 0} \left(\CJ^+_{AB}(\omega, z, \bz) + \CJ^-_{AB}(\omega, z, \bz) \right)
\ee
are the contributions from hard particles from $\ci^+$ and $\ci^-$, respectively. These currents may be traded for soft graviton insertions $\mathcal{N}_{AB}$ via the conservation of supertranslation charge, namely 
\be\label{eq:soft-eff-init}
\begin{split}
	 S_\KM &= -\frac{1}{16\pi^2} \int_{\ci^+_-} d^2z\, \CN_{AB} \widetilde{C}^{AB} + \CO(N^2), 
\end{split} 
\ee
where\footnote{Note that \cite{Kapec:2021eug} defines the shear as the $\CO(r)$ component of the transverse metric, which is the same as our definition. However, the news tensor differs from our $N_{AB}$ by a constant prefactor. Likewise, the Goldstone and memory modes $\CC,\CN$ defined in \cite{Kapec:2021eug} differs from ours by constant prefactors, all of which we will determine below.}
\begin{align}
\label{eq:C-N-KM}
	C_{AB} &= 2 \bigg( \p_A\p_B - \frac{1}{2} \g_{AB} \Box\bigg) \CC , \qquad \mathcal N_{AB} = 2 \bigg( \p_A\p_B - \frac{1}{2} \g_{AB} \Box\bigg) \CN .
\end{align}
For the memory shear, this is defined to be \cite{Kapec:2017gsg,Kapec:2021eug}\footnote{$C_{AB}$ has scaling dimension $\Delta =1$.}
\begin{align}
\begin{split}
	\widetilde{C}_{AB}(z,\bz) &= \int d^2w\, \frac{\CI_{AC}(z-w)\CI_{BD}(z-w)}{\g_{EF}(z-w)^E(z-w)^F } C^{CD}(w,\bw) , \qquad \CI_{AB}(x) = \g_{AB} - 2 \frac{x_A x_B}{\g_{CD}x^C x^D} .
\end{split}
\end{align}
We can now explicitly compute using the above equations 
\begin{align}\label{eq:CN-curly}
\begin{split}
	C_{zz} &= 2\p_z^2 \CC, \qquad \CN_{zz} = 2 \p_z^2 \CN , \qquad \CC_{z\bz} = \CN_{z\bz} = 0, \\
	\CI_{z\bz} &= 0 , \qquad \CI_{zz} = - \frac{\bz}{z},
\end{split}
\end{align}
which in turn implies
\begin{align}
	\widetilde{C}_{zz}(z,\bz) &= \int d^2w\, \frac{\bz-\bw}{2(z-w)^3} C_{\bw\bw}(w,\bw) .
\end{align}
Substituting these results into \eqref{eq:soft-eff-init}, we obtain
\be \label{eq:soft-eff-fin}
\begin{split}
	S_\KM &= -\frac{1}{4 \pi^2} \int_{\ci^+_-} d^2z\, d^2w\,  \bigg( \p_\bz^2 \CN(z,\bz)  \frac{\bz-\bw}{2(z-w)^3} \p_\bw^2 \CC(w,\bw)  + \cc \bigg) \\
	&= -\frac{1}{4 \pi^2} \int_{\ci^+_-} d^2z\, d^2w\,  \bigg( \frac{1}{4} \p_\bz^2 \CN(z,\bz) \p_\bw^2 \frac{\bz-\bw}{z-w} \p_w^2 \CC(w,\bw)  + \cc \bigg) \\
	&= -\frac{1}{4 \pi} \int_{\ci^+_-} d^2z  \,  \p_z^2 \CC \p_\bz^2 \CN ,
\end{split}
\ee
where we integrated by parts to simplify the expression, and in the last equality we used the second equality in \eqref{eq:useful}.

To relate this to our gauge-invariant on-shell action, we need to determine the respective relation between $\CC,\CN$ and our $C,N$. Comparing the first equation in \eqref{eq:CN-curly} with \eqref{eq:soft-shear} at $\ci^+_-$ (or $u = -\infty$), it becomes clear that
\begin{align}\label{eq:C-CC}
	\CC(z,\bz) = - C(z,\bz).
\end{align}
Next, to determine the relation between $\CN$ and $N$, we observe that $\CN_{AB}$ defined in \cite{Kapec:2021eug} is in fact
\begin{align}
	\CN_{AB} \equiv \CN_{AB}^+ - \CN_{AB}^- , \qquad \CN_{AB}^\pm \equiv \CN_{AB} \big|_{\ci^\pm_\mp}, 
\end{align}
and that
\begin{align}
	\langle \CN_{zz}^+ \cdots \rangle = \frac{2}{\kappa} \lim_{\omega \to 0} \omega \CS^{(0)+} , \qquad \kappa \equiv \sqrt{32\pi G_N},
\end{align}
where $\CS^{(0)+}$ is the Weinberg leading soft graviton factor for an outgoing positive helicity graviton. Noting that $\CN_{AB}$ obeys the analogous matching condition \eqref{eq:matching}, we obtain
\begin{align}\label{eq:N-eqn1}
	\langle \CN_{zz} \cdots \rangle = 2 \langle \CN_{zz}^+ \cdots \rangle = \frac{4}{\kappa} \lim_{\omega \to 0} \omega \CS^{(0)+} . 
\end{align}
On the other hand, we adopted the conventions of $N$ used in \cite{He:2014laa} with $\g_{z\bz} \to 1$. Thus, we have
\begin{align}\label{eq:N-eqn2}
	\langle N_{zz} \cdots \rangle = \langle \p_z^2 N \cdots \rangle = - \frac{\kappa}{8\pi } \lim_{\omega \to 0} \omega \CS^{(0) +} .
\end{align}
Comparing \eqref{eq:N-eqn1} with \eqref{eq:N-eqn2} and using \eqref{eq:CN-curly}, we get 
\begin{align}\label{eq:N-CN}
	\CN = - \frac{16\pi}{\kappa^2} N =  - \frac{1}{2G_N} N.
\end{align}
Substituting \eqref{eq:C-CC} and \eqref{eq:N-CN} into \eqref{eq:soft-eff-fin}, we get
\begin{align}
\label{eq:SKMfin}
	S_\KM &= - \frac{1}{8\pi G_N} \int_{\ci^+_-} d^2z\, \p_z^2 C \p_\bz^2 N ,
\end{align}
which is precisely the gauge-invariant on-shell action \eqref{eq:S-os-total}.

We conclude by remarking that in Section \ref{sec:bdry-act}, we obtained \eqref{eq:S-os-total} as an on-shell boundary action in \textit{pure} gravity. As discussed in Section \ref{sec:modular-Hamiltonian}, such a term may be non-vanishing even in the absence of hard particles (see also \cite{Donnay:2018neh,Chen:2023tvj,Chen:2024kuq}). Evaluated in a state that diagonalizes $C$, this agrees with the classical gravity computation subject to the boundary condition \eqref{eq:deltaC}, while evaluated in a state that diagonalizes $N$ instead, we obtain the infrared (Faddeev-Kulish) dressings in gravity \cite{Kulish:1970ut}. A dressing of the form \eqref{eq:SKMfin}, where both $N$ and $C$ are operators, has also been proposed in \cite{Freidel:2022skz} as the leading term in a series of terms pairing increasingly subleading memory modes with their corresponding Goldstone partners.

\section{Summary and future directions}
\label{sec:discussion}

In this work we have computed two on-shell actions associated with solutions to the Einstein equation in asymptotically flat spacetimes. The first solution reduces to Minkowski space in the limit $r \rightarrow \infty$ and is parameterized by an asymptotic shear profile including only a zero (Goldstone) mode $C(z,\bz)$ and a memory (soft) mode $N(z,\bz) \theta(u - u_0)$. The second solution is obtained from the former by a linearized diffeomorphism that eliminates the memory mode at the expense of turning on an $\mathcal{O}(1)$ gravitational potential at large $r$ localized in the retarded time $u$. We adopted boundary conditions that fix $C$ and are therefore preserved by the diffeomorphism considered.  We found that the on-shell actions differ by a corner term living on a codimension-2 surface at infinity. We have shown that this transformation is directly inherited from the transformation of the symplectic potential under field-dependent diffeomorphisms \cite{Donnelly:2016auv}. A diffeomorphism-invariant action was then obtained through the addition of an appropriate corner term. After adding the contributions from $\ci^{\pm}$, we found that the resulting diffeomorphism invariant boundary action agrees with the soft supertranslation charge, the Verlinde-Zurek action used in \cite{Verlinde:2022hhs} to compute quantum spacetime fluctuations in causal diamonds, and the interaction term in the soft effective action of Kapec and Mitra \cite{Kapec:2021eug}.  

There are many avenues for future research. Many of the outstanding puzzles were already outlined in Section \ref{sec:modular-Hamiltonian}. We will hence conclude by discussing a further aspect of our analysis that remains to be better understood, namely the relation between infrared divergences in gravity and the infrared action computed herein. We have for simplicity restricted our analysis here to linear order in $N$. Of course, the computation should go through beyond this approximation, and particularly interesting are the terms quadratic in $N$. These can already be included in the computation of the boundary action associated with the memory metric \eqref{eq:Mink-vacua}, \eqref{eq:soft-shear} starting from the expression for the memory symplectic potential \eqref{eq:soft-symp}. The coefficient of such terms is readily seen to be divergent, upon trading $m_B$ for gravitational degrees of freedom via the constraint \eqref{eq:mB-constraint}.  On the other hand, the soft effective action of \cite{Kapec:2021eug} also contains an infrared divergent term which is responsible for the soft S-matrix in gravity \cite{Weinberg:1965aa}. This takes the form
\be 
\label{eq:div-action}
S_{\rm div}^{\rm IR} = \log\frac{\Lambda_{\rm UV}}{\Lambda_{\rm IR}} \frac{G_N}{(2\pi)^2} \int d^2z \, \mathcal{N}_{zz} \mathcal{N}_{\bz\bz} =  \log\frac{\Lambda_{\rm UV}}{\Lambda_{\rm IR}} \frac{1}{(2\pi)^2 G_N} \int d^2z \, \p_z^2 N \p_{\bz}^2 N.
\ee
This term may be recovered from the corresponding term in the on-shell action subject to an identification of the divergences which takes the schematic form
\be 
\frac{1}{\pi} \log \frac{\Lambda_{UV}}{\Lambda_{IR}} \propto  \int du \int^u du' \, \delta(u' - u_0)^2.
\ee
This relation may be an important hint in the quest for the appropriate boundary conditions needed to render the path integral in gravity infrared finite. We leave a better understanding of  these ideas to future work.

\section*{Acknowledgements}

We would like to thank Laurent Freidel, Prahar Mitra, and Andrew Strominger for enlightening discussions, and especially Laurent Freidel for comments on a draft.  T.H. and
K.Z. are supported by the Heising-Simons Foundation
“Observational Signatures of Quantum Gravity” collaboration grant 2021-2817, the U.S. Department of Energy,
Office of Science, Office of High Energy Physics, under
Award No. DE-SC0011632, and the Walter Burke Institute for Theoretical Physics. A.R. has been supported in part by a Heising-Simons postdoctoral fellowship grant 2021-2817 and a Marie Curie fellowship grant agreement 101063234 at the University of Amsterdam, as well as the Simons Foundation through the Emmy Noether Fellows Program at
Perimeter Institute (1034867, Dittrich). A.R. is grateful for the hospitality of Perimeter Institute where part of this work was carried out. Research at Perimeter Institute is supported in part by the Government of Canada through the Department of Innovation, Science and Economic Development and by the Province of Ontario through the Ministry of Colleges and Universities. The work of K.Z. is also supported by a Simons Investigator award.

\appendix

\section{Asymptotically flat spacetime metrics}\label{app:AFS}

In this appendix, we derive \eqref{eq:Mink-vacua} by computing the general form of an asymptotically flat metric satisfying the Bondi gauge condition \eqref{eq:bondi}, which we reproduce here for convenience:
\begin{align}\label{bondi-gauge0}
	g_{rr} = g_{rA} = 0 , \qquad \p_r \sqrt{ \det \big( r^{-2} g_{AB} \big) } = 0 .
\end{align}
To this end, we first write the AFS metric \eqref{eq:afs-met} in planar null coordinates, including subleading terms, as
\begin{align}\label{bondi-010}
\begin{split}
	ds^2 &= - 2 du\,dr + 2r^2 \,dz\,d\bz \\
	&\qquad + \bigg[ \frac{2m_B}{r} + \frac{1}{r^2}g_{uu}^{\2} \bigg] du^2 + 2 \bigg[ \frac{1}{r}g^\1_{ur} + \frac{1}{r^2} g_{ur}^\2 \bigg] du\,dr \\
	&\qquad + 2 \bigg[ \bigg( g_{uz}^\0 + \frac{1}{r}g_{uz}^\1  \bigg)  du\,dz + \cc \bigg] + \Big[ \big( r C_{zz} + g_{zz}^\0  \big) dz^2 + \cc \Big] \\
	&\qquad + 2 \Big[ r g_{z\bz}^{(-1)} + g_{z\bz}^\0  \Big] dz\,d\bz .
\end{split}
\end{align}
We begin by requiring the determinant condition in \eqref{bondi-gauge0} holds to $\CO(r^{-2})$. This yields
\begin{align}
	\p_r \sqrt{ \det \big( r^{-2} g_{AB} \big) } = 0 \quad\implies\quad \boxed{ g_{z\bz}^{(-1)} = 0 , \quad g_{z\bz}^{\0} = \frac{1}{2} C_{zz}C^{zz}. }
\end{align}
Next, we solve for all components of the vacuum Einstein equation, which we denote as $\mee_{\mu\nu}$, order by order in large $r$. Both $\mee_{ur}$ and $\mee_{rr}$ automatically vanish at $\mathcal{O}(r^{-2})$:
\begin{align}
\begin{split}
	\mee_{ur} &= \CO(r^{-3})  , \qquad \mee_{rr} =  \CO(r^{-3})  .
\end{split}
\end{align}
Requiring the $\CO(r^0)$ component of $\mee_{z\bz}$ to vanish yields
\begin{align}
	\boxed{g_{ur}^\1 = 0.}
\end{align}
Substituting this back into $\mee_{rz}$ yields at $\CO(r^{-2})$
\begin{align}
\begin{split}
	\boxed{g_{uz}^\0 = \frac{1}{2} \p_\bz C_{zz} .}
\end{split}
\end{align}
Substituting these into $\mee_{uu}$, we find at $\mathcal{O}(r^{-2})$
\begin{align}\label{mB-constraint0}
\begin{split}
	 \boxed{ \p_u m_B = \frac{1}{4} (\p_z^2 N^{zz} + \cc ) - \frac{1}{4} N_{zz} N^{zz}, }
\end{split}
\end{align}
where $N_{zz} \equiv \p_u C_{zz}$ is the news tensor.

Going back to $\mee_{z\bz}$, we find that imposing $\mee_{z\bz} = \CO(r^{-2})$ requires
\begin{align}
	\boxed{ g_{ur}^\2 = \frac{1}{8} C_{zz}C^{zz}. }
\end{align}
Finally, substituting all the equations from above into $\mee_{zz}$ and requiring $\mee_{zz} = \CO(r^{-2})$, we get the condition
\begin{align}
	\boxed{ g_{zz}^\0 = 0 .}
\end{align}
Imposing the vacuum Einstein equation to further subleading order gives constraints on the subleading components. Thus, AFS metrics have the form at leading order in large-$r$ expansion
\begin{align}
\begin{split}
	ds^2 &= - 2 du\,dr + 2r^2 \,dz\,d\bz \\
	&\qquad +  \frac{2m_B}{r}  du^2 +  \frac{1}{4r^2} C_{zz}C^{zz}  du\,dr +  C_{zz}C^{zz} \, dz\,d\bz \\
	&\qquad + \Big[ \p_\bz C_{zz}   du\,dz + \cc \Big] + \Big[ r C_{zz}  \,dz^2 + \cc \Big] + \cdots,
\end{split}
\end{align}
which is precisely \eqref{eq:Mink-vacua}.
We note that in our analysis we assumed the large-$r$ expansions to contain no logarithmic terms. This assumption does not affect our analysis of the leading soft sector, but may need to be lifted in generalizations that include subleading memory effects.

\section{Gibbons-Hawking-York boundary term}
\label{app:bdy-terms}

In this appendix we start with a general discussion of the variational principle. We then review the Gibbons-Hawking-York (GHY) boundary term, as well as its relation to the on-shell action. We will derive the symplectic potential \eqref{eq:symp-pot} in Appendix~\ref{app:symp-pot}, and review the standard derivation of the GHY boundary term in Appendix~\ref{sec:bdry-on-shell}. Because our derivations work in any coordinate system, we will adopt abstract tensor indices throughout this section.

While our derivation of the on-shell action is similar in spirit to the standard one reviewed in this appendix, we emphasize that the formula for the GHY boundary action \eqref{GH-term} cannot be directly imported in the main text. One reason for this is that the boundary in our case is a null hypersurface as opposed to a timelike surface. Furthermore, the standard GHY boundary conditions cannot be imposed in the presence of radiation. Our analysis in the body of the paper will therefore only rely on the symplectic potential \eqref{theta-def} from this appendix.

\subsection{Deriving the symplectic potential}\label{app:symp-pot}

We will construct the symplectic potential via the covariant phase space formalism; for a recent explanation of the covariant phase space formalism, see \cite{Compere:2018aar,Harlow:2019yfa, He:2020ifr, Capone:2023roc}. Including matter, the EH action is given by\footnote{There is an additional boundary action term that we should include, but we will examine this term for the particular case of a timelike boundary with Dirichlet boundary conditions in Appendix~\ref{sec:bdry-on-shell}.}
\begin{align}\label{eq:action}
\begin{split}
	S_\EH &=  \frac{1}{16\pi G_N} \int d^4x\, \sqrt{-g} R + \int d^4 x\,L_M [ \Phi,g] \\
	&= \frac{1}{16\pi G_N} \int d^4x\, \sqrt{-g} g^{ab} R_{ab} + \int d^4 x\, L_M[\Phi,g],
\end{split}
\end{align}
where $g = \det g_{ab}$, $R$ is the Ricci scalar, and $L_M$ is the matter Lagrangian. In order to determine the symplectic potential density, we need to determine the boundary term that arises when we vary the action (i.e., the term not proportional to the equations of motion). The variation of $S_\EH$ yields
\begin{align}\label{eq:delta-S}
\begin{split}
	\delta S_{\EH} &= \frac{1}{16\pi G_N}\int d^4x \big( \sqrt{-g} g^{ab} \delta R_{ab} + \sqrt{-g} R_{ab} \delta g^{ab} + R \delta\sqrt{-g} \big) + \int d^4 x\, \delta L_M[\Phi,g] .
\end{split}
\end{align}
Note that we are defining $\delta g^{ab}$ as the variation of the inverse metric $g^{ab}$, so that
\begin{align}
	g_{ab} g^{bc} = \delta_a^c \quad\implies\quad \delta g_{ab} = -g_{ac}  \delta g^{cd} g_{db} .
\end{align}
To determine what $\delta R_{ab}$ is, we first recall
\begin{align}\label{eq:riemann}
	R^c{}_{adb} = \p_d \G^c_{ba} + \G^c_{de} \G^e_{ba} - (b \leftrightarrow d).
\end{align}
The variation of the Riemann tensor can thus be obtained by examining the variation of $\delta \G^c_{ba}$. By a straightforward computation, we find
\begin{align}
	\delta R^c{}_{adb} = \nabla_d(\delta\G^c_{ba}) - \nabla_b(\delta \G^c_{da} ),
\end{align}
and so the first term in the parentheses of \eqref{eq:delta-S} becomes
\begin{align}\label{delta1a}
\begin{split}
	\sqrt{-g} g^{ab} \delta R_{ab} = \sqrt{-g} g^{ab} \big( \nabla_c \delta\Gamma^c_{ba} - \nabla_b \delta \G^c_{ca} \big) .
\end{split}
\end{align}
To further simplify this, we recall
\begin{align}\label{Gamma-def}
	\G^c_{ab} = \frac{1}{2} g^{cd} \big( \p_a g_{bd} + \p_b g_{da} - \p_d g_{ab} \big).
\end{align}
We can directly vary this and obtain the equality\footnote{There is also a cleaner way of obtaining \eqref{eq:delta-Gamma-fin} by observing $\delta \G^c_{ab}$ is a tensor. Therefore we can work in Riemann flat coordinates and then simply covariantize the expression.}
\begin{align}\label{eq:delta-Gamma-fin}
	\delta\G^c_{ab} = \frac{1}{2} g^{cd}\big( \nabla_a \delta g_{bd} + \nabla_b \delta g_{da} - \nabla_d \delta g_{ab} \big).
\end{align}
Substituting this into \eqref{delta1a}, the first term in \eqref{eq:delta-S} becomes
\begin{align}\label{delta1}
\begin{split}
	\sqrt{-g} g^{ab} \delta R_{ab} &= \sqrt{-g} \nabla_c \big( g^{cd} \nabla^b \delta g_{bd} - g^{ab} \nabla^c \delta g_{ab} \big) .
\end{split}
\end{align}
Note that this is a total derivative term, and so will contribute to the symplectic potential.

Next, to evaluate the $R \delta \sqrt{-g}$ term in the integral in \eqref{eq:delta-S}, we recall the identity
\begin{align}\label{delta3}
\begin{split}
	\delta \sqrt{-g} =  - \frac{1}{2}\sqrt{-g} g_{ab} \delta g^{ab}.
\end{split}
\end{align}
Substituting \eqref{delta1} and \eqref{delta3} into \eqref{eq:delta-S} and rearranging the terms, we get 
\begin{align}\label{total-var}
\begin{split}
	\delta S_\EH &= \frac{1}{16\pi G_N}\int d^4x \sqrt{-g}    \nabla_c \big( g^{cd} \nabla^b \delta g_{bd} - g^{ab} \nabla^c \delta g_{ab} \big) \\
	&\qquad + \frac{1}{16\pi G_N}\int d^4x \sqrt{-g} \left[  R_{ab} - \frac{1}{2}R g_{ab} + \frac{16 \pi G_N}{\sqrt{-g}}\frac{\delta L_M}{\delta g^{ab}} \right] \delta g^{ab}  + \int d^4 x\, \frac{\delta L_M}{\delta \Phi} \delta \Phi.
\end{split}
\end{align}
The second line gives us the equations of motion, since if we identify the stress energy tensor to be
\begin{align}
	T_{ab} = -\frac{2}{\sqrt{-g}} \frac{\delta L_M}{\delta g^{ab}},
\end{align}
we arrive at the Einstein equation
\begin{align}\label{einstein}
	R_{ab} - \frac{1}{2}R g_{ab} = 8\pi G_N T_{ab} .
\end{align}
Similarly, we get (up to a boundary term that contributes to the symplectic potential) the equation of motion for the matter field by setting $\frac{\delta L_M}{\delta \Phi} = 0$. The first line, on the other hand, is a total boundary term, and results in the gravitational component of the symplectic potential
\begin{align}\label{theta-def}
\begin{split}
	\boxed{ \Th(g,\delta g) = \frac{1}{16\pi G_N}\int_{\p\CM} d\Sigma_a  \big( g^{ac} \nabla^b \delta g_{bc} - g^{cb} \nabla^a \delta g_{cb} \big) \equiv \frac{1}{16\pi G_N} \int_{\p\CM} d\Sigma_a \,\th^a  ,}
\end{split}
\end{align}
where $\p \CM$ is the boundary of the spacetime manifold. This is precisely the symplectic potential quoted in \eqref{eq:symp-pot}, as promised.

\subsection{Deriving the Gibbons-Hawking-York boundary action}
\label{sec:bdry-on-shell}

We now turn to discussing the boundary contribution to the action \eqref{eq:action}, which is fixed by requiring the variational principle to be well-defined subject to boundary conditions on the metric degrees of freedom. The case of Dirichlet boundary conditions, where the metric components tangential to the boundary are fixed, has been analyzed in \cite{York, Gibbons-Hawking}. Such conditions are typically imposed on timelike boundaries,\footnote{In the cases we will be interested in, the spacelike and null boundaries will be Cauchy surfaces where we shall instead specify initial data. } in which case the bulk metric decomposes as 
\be 
\label{eq:metric-dec}
	g_{ab} = \gamma_{ab} + n_{a}n_{b}, \qquad n_{a} n^{a} = 1, \qquad  \g_{ab} n^a = 0,
\ee
where $\g_{ab}$ is the induced metric on the boundary, and $n^a$ is the spacelike normal to the boundary.  From \eqref{eq:metric-dec}, we observe the useful identities
\begin{align}\label{useful-var}
	0 = \delta(n_an^a) = 2 n^a \delta n_a , \qquad 0 = \nabla_a(n^c n_c) = 2n^c \nabla_a n_c.
\end{align}
Using \eqref{eq:metric-dec}, the integrand of the gravitational symplectic potential from \eqref{theta-def} can be rewritten as\footnote{Recall $d\Sigma_a = d^3x \sqrt{-\gamma}n_a$ on the boundary $\p\CM$.}
\be \label{eq:theta-gamma}
\begin{split}
	 n_a \th^a \big|_{\p\CM} &= \big( n^c g^{ba} \nabla_a \delta g_{bc} - n^a g^{cb} \nabla_a \delta g_{cb} \big) \big|_{\p\CM}  \\
	&= \big( n^c \g^{ba} \nabla_a \delta g_{bc} - n^a \g^{cb} \nabla_a \delta g_{cb} \big)\big|_{\p\CM} , 
\end{split}
\ee
where in the second line we expanded $g^{ab}$ using \eqref{eq:metric-dec} and then canceled terms.

Next, the authors of \cite{York, Gibbons-Hawking} proceeded by imposing Dirichlet boundary conditions, which means they only considered the class of variations that kept $\p\CM$ fixed. This means that the metric and tangential derivatives of the metric remain fixed on $\p\CM$, so that 
\begin{align}\label{gamma-var}
\begin{split}
	\delta g_{ab} \big|_{\p\CM} &= 0 \quad\implies\quad \delta \g_{ab}\big|_{\p\CM} = \delta n_a \big|_{\p\CM}  = 0, 
\end{split}
\end{align}
and
\begin{align}\label{dn-var}
\begin{split}
	\g^{ab}\p_b \delta g_{cd} \big|_{\p\CM} = 0 \quad &\implies\quad \g^{ab}\p_b \delta (n_cn_d) \big|_{\p\CM} = 0 \quad\implies\quad \g^{ab}\p_b \delta n_c \big|_{\p\CM} = 0,
\end{split}
\end{align}
where the first implication in \eqref{dn-var} used \eqref{gamma-var}. Note that we further have
\begin{align}\label{dn-var2}
	\g^{ab} \nabla_a \delta g_{cd} \big|_{\p\CM} = \g^{ab} \big( \p_a \delta g_{cd} - \G^e_{ac} \delta g_{ed} - \G^e_{ad} \delta g_{ce} \big) \big|_{\p\CM} = 0,
\end{align}
where the last equality follows from \eqref{gamma-var} and the first equality in \eqref{dn-var}. Substituting this into \eqref{eq:theta-gamma} yields
\begin{align}\label{n-theta2}
	n_a \th^a \big|_{\p\CM} = - n^a\g^{cb} \nabla_a \delta g_{cb} \big|_{\p\CM}.
\end{align}
We will show that this term is proportional to the extrinsic curvature below.

The extrinsic curvature is defined on the boundary manifold $\p\CM$ to be
\begin{align}\label{K-def}
\begin{split}
	K &\equiv g^{ab} \nabla_a n_b \big|_{\p\CM} = \g^{ab} \nabla_a n_b \big|_{\p\CM} = \g^{ab} \big(\p_a n_c - \G_{ab}^c n_c \big) \big|_{\p\CM} ,
\end{split}
\end{align}
where in the second equality we used \eqref{useful-var}. Using \eqref{gamma-var}, the definition of the Christoffel symbol \eqref{Gamma-def} implies
\begin{align}\label{var-Gamma-bdy}
\begin{split}
	\delta \G^c_{ab} \big|_{\p\CM} &= \frac{1}{2} g^{cd} \big( \p_a \delta g_{bd} + \p_b \delta g_{da} - \p_d \delta g_{ab} \big) \big|_{\p\CM} .
\end{split}
\end{align}
Thus, the variation of \eqref{K-def} is
\begin{align}
\begin{split}
	\delta K &= - \g^{ab} \delta \G^c_{ab} n_c \big|_{\p\CM} \\
	&= - \frac{1}{2} \g^{ab} n_c g^{cd} \big( \p_a \delta g_{bd} + \p_b \delta g_{da} - \p_d \delta g_{ab} \big) \big|_{\p\CM} \\
	&= \frac{1}{2} \g^{ab} n^c \p_c \delta g_{ab} \big|_{\p\CM},
\end{split}
\end{align}
where in the first equality we used \eqref{gamma-var} and \eqref{dn-var}; in the second equality we used \eqref{var-Gamma-bdy}; and in the third equality we used the first equality in \eqref{dn-var}. Substituting this into \eqref{n-theta2}, we have
\begin{align}
\begin{split}
	n_a \th^a \big|_{\p\CM} &= - 2 \delta K \big|_{\p\CM} .
\end{split}
\end{align}
Therefore, we see that the variation of the bulk EH term, i.e., the symplectic potential, is given by
\begin{align}
	\delta S_{\EH} \equiv \Th(g,\delta g) = -\frac{1}{8\pi G_N} \int_{\p\CM} d^3 x\,\sqrt{-\g}\, \delta K,
\end{align}
where we used the fact $d\Sigma_a = d^3x \sqrt{-\g} n_a$. To ensure the variational principle is satisfied, we need to add a boundary term, namely the GHY boundary action, whose variation cancels $\delta S_\EH$. It is obvious then that this is given by
\begin{align}\label{GH-term}
\begin{split}
	S_\text{GHY} = \frac{1}{8\pi G_N} \int_{\p\CM} d^3x\, \sqrt{-\g} K .
\end{split}
\end{align}

From the above computation, we see that the GHY boundary action, or the gravitational on-shell action, is indeed a boundary term. For the vacuum shockwave and memory spacetimes that we are interested in, the boundary $\p\CM$ is null and not timelike. Therefore, the standard computation of the GHY boundary action given above will not directly apply. Nevertheless, we will evaluate the gravitational on-shell action for the vacuum shockwave and memory spacetimes in Section~\ref{sec:bdry-act}. Interestingly, we shall see that for these two cases, the on-shell action is not just a codimension-1 boundary term, but a codimension-2 \emph{corner term} (cf. \eqref{eq:mem-bdy-action} and \eqref{eq:bdry-shock}). By adding an additional corner contribution, we are able to get an on-shell action that is invariant under the linearized diffeomorphisms considered in this paper (cf. \eqref{eq:S-os-total}).

\section{Further details concerning shockwave metrics}
\label{app:sym}

In this appendix, we want to explore some further properties of the shockwave metric. In Appendix~\ref{app:shockwave-met}, we will derive a more general class of shockwave metrics than that given in \eqref{eq:shock-met2} that satisfies the vacuum Einstein equation. As we will see, \eqref{eq:shock-met2} is a special case obtained from the solution derived here by setting $C_{zz} = -2\p_z^2C$ and linearizing in $g_{ur}^{(0)}$. In Appendix~\ref{app:diff-action-symp-pot}, we will then perform an explicit check that the memory symplectic potential transforms under a linearized diffeomorphism into the shockwave symplectic potential via \eqref{eq:diff-transform-pot}.

\subsection{Deriving shockwave metrics}\label{app:shockwave-met}

We want to perform the analogous calculation we did in Appendix~\ref{app:AFS}, except we will now allow the metric to include shockwaves, i.e. a $\CO(r^0)$ contribution to $g_{uu}$ proportional to $\delta(u-u_0)$. As we shall see, to ensure that the vacuum Einstein equation, denoted as $\mee_{\mu\nu}$, is obeyed, we need to include additional divergent $g_{ur}$ and $g_{uz}$ terms to the AFS metric \eqref{bondi-010}. As in Appendix~\ref{app:AFS}, we first write the shockwave metric, including subleading corrections, in the general form
\begin{align}\label{shock-01}
\begin{split}
	ds^2 &= - 2 du\,dr + 2r^2 \,dz\,d\bz \\
	&\qquad +  \bigg[ \a(u,z,\bz) + \frac{2m_B'}{r} \bigg]  du^2 +  2 \bigg[ g_{ur}^\0 + \frac{1}{r} g_{ur}^\1 + \frac{1}{r^2} g_{ur}^\2 \bigg]  du\,dr  \\
	&\qquad + 2 \bigg[ \bigg( rg_{uz}^{(-1)} + g_{uz}^\0 + \frac{1}{r}g_{uz}^\1  \bigg)  du\,dz + \cc \bigg] + \Big[ \big( r C_{zz} + g_{zz}^\0 \big)  \,dz^2 + \cc \Big] \\
	&\qquad + 2 \Big[ r g_{z\bz}^{(-1)} +  g_{z\bz}^\0 \Big] \, dz\,d\bz ,
\end{split}
\end{align}
where $m_B'$ is the Bondi mass associated to the shockwave metric. 

As before, we impose Bondi gauge \eqref{bondi-gauge0} to $\CO(r^{-2})$, so that
\begin{align}
	\p_r \sqrt{ \det \big( r^{-2} g_{AB} \big)  } = 0 \quad\implies\quad \boxed{ g_{z\bz}^{(-1)} = 0 , \quad g_{z\bz}^{\0} = \frac{1}{2} C_{zz}C^{zz}. }
\end{align}
Note that $\mee_{rr}$ automatically vanishes at $\CO(r^{-2})$, namely
\begin{align}
\begin{split}
	\mee_{rr} =  \CO(r^{-3})  .
\end{split}
\end{align}
We next compute the $\CO(r^{-1})$ component of $\mee_{rz}$, which gives us the condition
\begin{align}\label{uz-ur}
	 \boxed{ g_{uz}^{(-1)} = \p_z g_{ur}^\0 . }
\end{align}
Substituting this back and requiring that $\mee_{ur}$ vanishes at $\CO(r^{-2})$ implies the shockwave parameter $\a$ is given by 
\begin{align}\label{alpha-constraint0}
\begin{split}
	\boxed{ \a = \Big[ 2 \p_z g_{ur}^\0 \p_\bz g_{ur}^\0 - 2 \big(g_{ur}^\0 - 1 \big) \p_z\p_\bz g_{ur}^\0 \Big] . }
\end{split}
\end{align}
Substituting these back into $\mee_{z\bz}$ and requiring that it vanishes at $\CO(r^{0})$ yields\footnote{More generally, we can have $g_{ur}^\1 = \Psi_1(z,\bz)$ for an arbitrary function on the sphere, but we consider the $\Psi_1=0$ case for simplicity.}
\begin{align}
	\boxed{g_{ur}^\1 = 0. }
\end{align}
Further evaluating $\mee_{rz}$ and requiring it to vanish at $\CO(r^{-2})$, we obtain
\begin{align}
	\boxed{ g_{uz}^\0 = \frac{1}{2} \Big[ \p_\bz g_{ur}^\0 C_{zz} - \big( g_{ur}^\0 - 1 \big) \p_\bz C_{zz} \Big]  . }
\end{align}
Turning to $\mee_{zz}$ and requiring it to vanish at $\CO(r^{-1})$, we obtain\footnote{More generally, we can have $g_{zz}^\0 = \Psi_2(z,\bz)$ for an arbitrary function on the sphere, but we consider the $\Psi_2=0$ case for simplicity.}
\begin{align}
	\boxed{g_{zz}^\0 = 0 .}
\end{align}
Finally, returning back to $\mee_{z\bz}$, but this time requiring it to vanish at $\CO(r^{-1})$, yields the constraint\footnote{More generally, we can have $g_{ur}^{\2} = \frac{1}{8}C_{zz}C^{zz} + \Psi_3(z,\bz)(g_{ur}^\0 -1 )$, but we set $\Psi_3 = 0$ for simplicity.}
\begin{align}
	\boxed{ g_{ur}^\2 = -\frac{1}{8}  C_{zz}C^{zz} \big(g_{ur}^\0 - 1 \big)  = \frac{1}{8} C_{zz} C^{zz} + \cdots ,}
\end{align}
where $\cdots$ indicate cubic terms in the fields that we are dropping in our analysis. Finally, evaluating $\mee_{uu}$ and requiring $\mee_{uu} = \CO(r^{-3})$ yields the constraint equation for $m_B'$, which in general is rather complicated. Likewise, evaluating $\mee_{uz}$ and requiring it to vanish at $\CO(r^{-2})$ fixes the subleading term $g_{uz}^\1$. As we will not require either constraint equation, we will not write them.

To summarize, the most general shockwave metric (ignoring the caveats mentioned in the footnotes in this subsection) satisfying the vacuum Einstein equation is to leading order in a large-$r$ expansion (and to quadratic order in the fields) given by
\begin{align}\label{shock-02}
\begin{split}
	ds^2 &= - 2 du\,dr + 2r^2 \,dz\,d\bz \\
	&\qquad +  \bigg[ \a(u,z,\bz) + \frac{2m_B'}{r} \bigg]  du^2 +  \bigg[ 2 g_{ur}^\0  + \frac{1}{4r^2} C_{zz}C^{zz}  \bigg]  du\,dr +  C_{zz}C^{zz} \, dz\,d\bz \\
	&\qquad + \bigg[ \bigg( 2 r \p_z g_{ur}^\0 +  \p_\bz g_{ur}^\0 C_{zz} - \big( g_{ur}^\0 - 1 \big) \p_\bz C_{zz}    \bigg)  du\,dz + \cc \bigg] + \bigg[ r C_{zz}  \,dz^2 + \cc \bigg] ,
\end{split}
\end{align}
with $\a$ obeying the constraint equation \eqref{alpha-constraint0}, and $m_B'$ obeying the constraint equation arising from solving the $uu$-component of the vacuum EE. This is a more general class of shockwaves than the one we considered. To restrict to the shockwave metric in \eqref{eq:shock-met2}, it is straightforward to check that we simply need to make the replacement
\begin{align}
\begin{split}
	g_{ur}^\0(u,z,\bz) = - \frac{1}{2} N(z,\bz)\delta(u-u_0) , \qquad C_{zz} = -2\p_z^2 C,
\end{split}
\end{align}
and then drop all terms quadratic in $N$ in \eqref{shock-02}.

\subsection{Transforming the symplectic potential under linearized diffeomorphisms}
\label{app:diff-action-symp-pot}

We now want to explicitly verify that \eqref{eq:shock-symp} can be directly obtained from \eqref{eq:symp-pot-coords} by applying the formula \eqref{eq:diff-transform-pot} for the special case of a linearized diffeomorphism generated by the vector $\xi^a$ given in \eqref{eq:full-diffeo}. Let the pullback $\phi^*$ in \eqref{eq:diff-transform-pot} take us from the memory metric to the shockwave metric. Note that the pullback acts on the symplectic potential density of the memory metric on the right-hand-side, and hence maps the memory metric determinant $\sqrt{-g_\mem}$ to the shockwave metric determinant $\sqrt{-g_\shock}$. Thus, the linearized version of \eqref{eq:diff-transform-pot} that we want to verify is 
\begin{align}\label{eq:transform}
\begin{split}
	\sqrt{-g_\shock} \th_\shock^\mu = \sqrt{-g_{\shock}} \th^\mu_\mem + \sqrt{-g_\shock}\CL_\xi\th^\mu_\mem  + \sqrt{-g_\shock} \th^\mu(\delta, 2\nabla_{(a} \delta \xi_{b)} ) ,
\end{split}
\end{align}
where $\sqrt{-g_\shock}$ is the shockwave metric determinant factor given in \eqref{eq:shock-det}, $\nabla_{(a} \delta \xi_{b)} \equiv \frac{1}{2} (\nabla_a \delta \xi_b + \nabla_b \delta \xi_a)$ is the symmetrized spacetime 2-form, and $\th^\mu(\delta, 2\nabla_{(a} \delta \xi_{b)} )$ is the symplectic potential density given in \eqref{eq:symp-pot} with $\delta g \to 2\nabla_{(a} \delta \xi_{b)} $, i.e.,
\begin{align}\label{eq:extra}
\begin{split}
	\th^\mu(g, 2\nabla_{(a} \delta \xi_{b)} ) &= 2 g^{\sigma\mu} \nabla^\nu \nabla_{(\mu} \delta \xi_{\nu)} - 2 g^{\mu\nu} \nabla^\sigma \nabla_{(\mu} \delta \xi_{\nu)} .
\end{split}
\end{align}

To confirm \eqref{eq:transform}, we first recall that the memory metric symplectic potential current density is given by \eqref{eq:AF-sp} and \eqref{eq:soft-symp} to be
\begin{align}
\begin{split}
	\th^u_\mem &= \th^z_\mem = \CO(r^{-3}) \\
	\th^r_\mem &= \frac{1}{2r^2} \bigg[ 4 \delta m_B + 4\p_z^2\p_\bz^2 \delta C - 2\p_z^2\p_\bz^2 N \th(u-u_0) + \frac{1}{4} \delta \big( \p_z^2 N \p_\bz^2 N  \big) \delta(u-u_0)  \\
	&\qquad + \frac{1}{2} \delta \big(  \p_z^2 C \p_\bz^2 N  + \cc \big)\delta(u-u_0) - 2\big( \p_z^2 \delta C \p_\bz^2 N + \cc \big) \delta(u-u_0)  \bigg] + \CO(r^{-3}) .
\end{split}
\end{align}
Next, we straightforwardly compute to linear order in $N$ that
\begin{align}
\begin{split}
	\sqrt{-g_\shock} \CL_\xi \th^\mu_\mem = \CO(r^{-1}),
\end{split}
\end{align}
where we used \eqref{eq:mB-constraint} to get the above result. Hence we can ignore this term. Finally, we compute \eqref{eq:extra} to linear order in $N$ and bilinear in $C,N$ to be
\begin{align}\label{eq:theta-sym}
\begin{split}
	\th^u(g, 2\nabla_{(a} \delta \xi_{b)} ) &= -\frac{1}{r} \delta N \delta(u-u_0) \\
	\th^r(g, 2\nabla_{(a} \delta \xi_{b)} ) &= \frac{1}{2} \delta N \delta'(u-u_0) - \frac{1}{r} \p_z\p_\bz \delta N \delta(u-u_0) \\
	&\qquad + \frac{1}{r^2} \bigg[ -m_B \delta N \delta(u-u_0) + \p_z^2\p_\bz^2 \delta N  \th(u-u_0) \\
	&\qquad - \frac{1}{4} \big( 2\p_z\delta N \p_z\p_\bz^2 C + \p_z^2 \delta N \p_\bz^2 C + \p_z^2 N \p_\bz^2 \delta C  + \cc \big) \delta(u-u_0)  \bigg] \\
	\th^z(g, 2\nabla_{(a} \delta \xi_{b)} ) &= \frac{1}{2r^2} \p_\bz \delta N \delta(u-u_0)  ,
\end{split}
\end{align}
where we have implicitly used \eqref{eq:mB-constraint} to drop terms involving $\delta N \p_u m_B$ as they are $\CO(N^2)$. Substituting the above results into the right-hand-side of \eqref{eq:transform}, we recover precisely \eqref{eq:shock-theta} mulitiplied by the determinant factor $\sqrt{-g_\shock}$, thereby confirming \eqref{eq:transform}.

\bibliography{VZeffect-bib}{}

\providecommand{\href}[2]{#2}\begingroup\raggedright\begin{thebibliography}{10}

\bibitem{He:2023qha}
T.~He, A.-M. Raclariu, and K.~M. Zurek, ``{From shockwaves to the gravitational
  memory effect},'' \href{http://dx.doi.org/10.1007/JHEP01(2024)006}{{\em JHEP}
  {\bfseries 01} (2024) 006}, \href{http://arxiv.org/abs/2305.14411}{{\ttfamily
  arXiv:2305.14411 [hep-th]}}.

\bibitem{Verlinde:2022hhs}
E.~Verlinde and K.~M. Zurek, ``{Modular fluctuations from shockwave
  geometries},'' \href{http://dx.doi.org/10.1103/PhysRevD.106.106011}{{\em
  Phys. Rev. D} {\bfseries 106} no.~10, (2022) 106011},
  \href{http://arxiv.org/abs/2208.01059}{{\ttfamily arXiv:2208.01059
  [hep-th]}}.

\bibitem{Verlinde:1991iu}
H.~L. Verlinde and E.~P. Verlinde, ``{Scattering at Planckian energies},''
  \href{http://dx.doi.org/10.1016/0550-3213(92)90236-5}{{\em Nucl. Phys. B}
  {\bfseries 371} (1992) 246--268},
  \href{http://arxiv.org/abs/hep-th/9110017}{{\ttfamily arXiv:hep-th/9110017}}.

\bibitem{Liu:2021kay}
S.~Liu and B.~Yoshida, ``{Soft thermodynamics of gravitational shock wave},''
  \href{http://dx.doi.org/10.1103/PhysRevD.105.026003}{{\em Phys. Rev. D}
  {\bfseries 105} no.~2, (2022) 026003},
  \href{http://arxiv.org/abs/2104.13377}{{\ttfamily arXiv:2104.13377
  [hep-th]}}.

\bibitem{tHooft:1996rdg}
G.~'t~Hooft, ``{The Scattering matrix approach for the quantum black hole: An
  Overview},'' \href{http://dx.doi.org/10.1142/S0217751X96002145}{{\em Int. J.
  Mod. Phys. A} {\bfseries 11} (1996) 4623--4688},
  \href{http://arxiv.org/abs/gr-qc/9607022}{{\ttfamily arXiv:gr-qc/9607022}}.

\bibitem{Ball:2018prg}
A.~Ball, M.~Pate, A.-M. Raclariu, A.~Strominger, and R.~Venugopalan,
  ``{Measuring color memory in a color glass condensate at
  electron\textendash{}ion colliders},''
  \href{http://dx.doi.org/10.1016/j.aop.2019.04.010}{{\em Annals Phys.}
  {\bfseries 407} (2019) 15--28},
  \href{http://arxiv.org/abs/1805.12224}{{\ttfamily arXiv:1805.12224
  [hep-ph]}}.

\bibitem{Kapec:2021eug}
D.~Kapec and P.~Mitra, ``{Shadows and soft exchange in celestial CFT},''
  \href{http://dx.doi.org/10.1103/PhysRevD.105.026009}{{\em Phys. Rev. D}
  {\bfseries 105} no.~2, (2022) 026009},
  \href{http://arxiv.org/abs/2109.00073}{{\ttfamily arXiv:2109.00073
  [hep-th]}}.

\bibitem{He:2024ddb}
T.~He, P.~Mitra, A.~Sivaramakrishnan, and K.~M. Zurek, ``{On-shell derivation
  of the soft effective action in Abelian gauge theories},''
  \href{http://dx.doi.org/10.1103/PhysRevD.109.125016}{{\em Phys. Rev. D}
  {\bfseries 109} no.~12, (2024) 125016},
  \href{http://arxiv.org/abs/2403.14502}{{\ttfamily arXiv:2403.14502
  [hep-th]}}.

\bibitem{York}
J.~W. York, ``Role of conformal three-geometry in the dynamics of
  gravitation,'' \href{http://dx.doi.org/10.1103/PhysRevLett.28.1082}{{\em
  Phys. Rev. Lett.} {\bfseries 28} (Apr, 1972) 1082--1085}.
  \url{https://link.aps.org/doi/10.1103/PhysRevLett.28.1082}.

\bibitem{Gibbons-Hawking}
G.~W. Gibbons and S.~W. Hawking, ``Action integrals and partition functions in
  quantum gravity,'' \href{http://dx.doi.org/10.1103/PhysRevD.15.2752}{{\em
  Phys. Rev. D} {\bfseries 15} (May, 1977) 2752--2756}.
  \url{https://link.aps.org/doi/10.1103/PhysRevD.15.2752}.

\bibitem{Fabbrichesi:1993kz}
M.~Fabbrichesi, R.~Pettorino, G.~Veneziano, and G.~A. Vilkovisky, ``{Planckian
  energy scattering and surface terms in the gravitational action},''
  \href{http://dx.doi.org/10.1016/0550-3213(94)90361-1}{{\em Nucl. Phys. B}
  {\bfseries 419} (1994) 147--188},
  \href{http://arxiv.org/abs/hep-th/9309037}{{\ttfamily arXiv:hep-th/9309037}}.

\bibitem{Strominger:2014pwa}
A.~Strominger and A.~Zhiboedov, ``{Gravitational Memory, BMS Supertranslations
  and Soft Theorems},'' \href{http://dx.doi.org/10.1007/JHEP01(2016)086}{{\em
  JHEP} {\bfseries 01} (2016) 086},
\href{http://arxiv.org/abs/1411.5745}{{\ttfamily arXiv:1411.5745 [hep-th]}}.

\bibitem{Strominger:2013jfa}
A.~Strominger, ``{On BMS Invariance of Gravitational Scattering},''
  \href{http://dx.doi.org/10.1007/JHEP07(2014)152}{{\em JHEP} {\bfseries 07}
  (2014) 152},
\href{http://arxiv.org/abs/1312.2229}{{\ttfamily arXiv:1312.2229 [hep-th]}}.

\bibitem{Kapec:2022hih}
D.~Kapec, ``{Soft Particles and Infinite-Dimensional Geometry},''
  \href{http://arxiv.org/abs/2210.00606}{{\ttfamily arXiv:2210.00606
  [hep-th]}}.

\bibitem{Capone:2023roc}
F.~Capone, P.~Mitra, A.~Poole, and B.~Tomova, ``{Phase space renormalization
  and finite BMS charges in six dimensions},''
  \href{http://dx.doi.org/10.1007/JHEP11(2023)034}{{\em JHEP} {\bfseries 11}
  (2023) 034}, \href{http://arxiv.org/abs/2304.09330}{{\ttfamily
  arXiv:2304.09330 [hep-th]}}.

\bibitem{Strominger:2017zoo}
A.~Strominger, ``{Lectures on the Infrared Structure of Gravity and Gauge
  Theory},''
\href{http://arxiv.org/abs/1703.05448}{{\ttfamily arXiv:1703.05448 [hep-th]}}.

\bibitem{Donnay:2018neh}
L.~Donnay, A.~Puhm, and A.~Strominger, ``{Conformally Soft Photons and
  Gravitons},'' \href{http://dx.doi.org/10.1007/JHEP01(2019)184}{{\em JHEP}
  {\bfseries 01} (2019) 184},
\href{http://arxiv.org/abs/1810.05219}{{\ttfamily arXiv:1810.05219 [hep-th]}}.

\bibitem{Freidel:2022skz}
L.~Freidel, D.~Pranzetti, and A.-M. Raclariu, ``{A discrete basis for celestial
  holography},'' \href{http://arxiv.org/abs/2212.12469}{{\ttfamily
  arXiv:2212.12469 [hep-th]}}.

\bibitem{Donnelly:2016auv}
W.~Donnelly and L.~Freidel, ``{Local subsystems in gauge theory and gravity},''
  \href{http://dx.doi.org/10.1007/JHEP09(2016)102}{{\em JHEP} {\bfseries 09}
  (2016) 102}, \href{http://arxiv.org/abs/1601.04744}{{\ttfamily
  arXiv:1601.04744 [hep-th]}}.

\bibitem{Ruzziconi:2020wrb}
R.~Ruzziconi and C.~Zwikel, ``{Conservation and Integrability in
  Lower-Dimensional Gravity},''
  \href{http://dx.doi.org/10.1007/JHEP04(2021)034}{{\em JHEP} {\bfseries 04}
  (2021) 034}, \href{http://arxiv.org/abs/2012.03961}{{\ttfamily
  arXiv:2012.03961 [hep-th]}}.

\bibitem{Weinberg:1965aa}
S.~Weinberg, ``Infrared photons and gravitons,''
  \href{http://dx.doi.org/10.1103/PhysRev.140.B516}{{\em Physical Review}
  {\bfseries 140} no.~2B, (1965) B516--B524}.

\bibitem{Bondi:1962px}
H.~Bondi, M.~G.~J. van~der Burg, and A.~W.~K. Metzner, ``{Gravitational waves
  in general relativity. 7. Waves from axisymmetric isolated systems},''
  \href{http://dx.doi.org/10.1098/rspa.1962.0161}{{\em Proc. Roy. Soc. Lond. A}
  {\bfseries 269} (1962) 21--52}.

\bibitem{Sachs:1962zza}
R.~Sachs, ``{Asymptotic symmetries in gravitational theory},''
  \href{http://dx.doi.org/10.1103/PhysRev.128.2851}{{\em Phys. Rev.} {\bfseries
  128} (1962) 2851--2864}.

\bibitem{Barnich:2010eb}
G.~Barnich and C.~Troessaert, ``{Aspects of the BMS/CFT correspondence},''
  \href{http://dx.doi.org/10.1007/JHEP05(2010)062}{{\em JHEP} {\bfseries 05}
  (2010) 062}, \href{http://arxiv.org/abs/1001.1541}{{\ttfamily arXiv:1001.1541
  [hep-th]}}.

\bibitem{Tamburino:1966zz}
L.~A. Tamburino and J.~H. Winicour, ``{Gravitational Fields in Finite and
  Conformal Bondi Frames},''
  \href{http://dx.doi.org/10.1103/PhysRev.150.1039}{{\em Phys. Rev.} {\bfseries
  150} (1966) 1039--1053}.

\bibitem{He:2014laa}
T.~He, V.~Lysov, P.~Mitra, and A.~Strominger, ``{BMS supertranslations and
  Weinbergs soft graviton theorem},''
  \href{http://dx.doi.org/10.1007/JHEP05(2015)151}{{\em JHEP} {\bfseries 05}
  (2015) 151},
\href{http://arxiv.org/abs/1401.7026}{{\ttfamily arXiv:1401.7026 [hep-th]}}.

\bibitem{Compere:2016hzt}
G.~Comp\`ere and J.~Long, ``{Classical static final state of collapse with
  supertranslation memory},''
  \href{http://dx.doi.org/10.1088/0264-9381/33/19/195001}{{\em Class. Quant.
  Grav.} {\bfseries 33} no.~19, (2016) 195001},
  \href{http://arxiv.org/abs/1602.05197}{{\ttfamily arXiv:1602.05197 [gr-qc]}}.

\bibitem{Compere:2016jwb}
G.~Comp\`ere and J.~Long, ``{Vacua of the gravitational field},''
  \href{http://dx.doi.org/10.1007/JHEP07(2016)137}{{\em JHEP} {\bfseries 07}
  (2016) 137}, \href{http://arxiv.org/abs/1601.04958}{{\ttfamily
  arXiv:1601.04958 [hep-th]}}.

\bibitem{tHooft:1987vrq}
G.~'t~Hooft, ``{Graviton Dominance in Ultrahigh-Energy Scattering},''
  \href{http://dx.doi.org/10.1016/0370-2693(87)90159-6}{{\em Phys. Lett. B}
  {\bfseries 198} (1987) 61--63}.

\bibitem{Carroll:2004st}
S.~M. Carroll, \href{http://dx.doi.org/10.1017/9781108770385}{{\em {Spacetime
  and Geometry}: {An Introduction to General Relativity}}}.
\newblock Cambridge University Press, 7, 2019.

\bibitem{Barnich:2011mi}
G.~Barnich and C.~Troessaert, ``{BMS charge algebra},''
  \href{http://dx.doi.org/10.1007/JHEP12(2011)105}{{\em JHEP} {\bfseries 12}
  (2011) 105}, \href{http://arxiv.org/abs/1106.0213}{{\ttfamily arXiv:1106.0213
  [hep-th]}}.

\bibitem{Campiglia:2014yka}
M.~Campiglia and A.~Laddha, ``{Asymptotic symmetries and subleading soft
  graviton theorem},'' \href{http://dx.doi.org/10.1103/PhysRevD.90.124028}{{\em
  Phys. Rev.} {\bfseries D90} no.~12, (2014) 124028},
\href{http://arxiv.org/abs/1408.2228}{{\ttfamily arXiv:1408.2228 [hep-th]}}.

\bibitem{Freidel:2021fxf}
L.~Freidel, R.~Oliveri, D.~Pranzetti, and S.~Speziale, ``{The Weyl BMS group
  and Einstein\textquoteright{}s equations},''
  \href{http://dx.doi.org/10.1007/JHEP07(2021)170}{{\em JHEP} {\bfseries 07}
  (2021) 170}, \href{http://arxiv.org/abs/2104.05793}{{\ttfamily
  arXiv:2104.05793 [hep-th]}}.

\bibitem{Geiller:2022vto}
M.~Geiller and C.~Zwikel, ``{The partial Bondi gauge: Further enlarging the
  asymptotic structure of gravity},''
  \href{http://dx.doi.org/10.21468/SciPostPhys.13.5.108}{{\em SciPost Phys.}
  {\bfseries 13} (2022) 108}, \href{http://arxiv.org/abs/2205.11401}{{\ttfamily
  arXiv:2205.11401 [hep-th]}}.

\bibitem{Geiller:2024amx}
M.~Geiller and C.~Zwikel, ``{The partial Bondi gauge: Gauge fixings and
  asymptotic charges},''
  \href{http://dx.doi.org/10.21468/SciPostPhys.16.3.076}{{\em SciPost Phys.}
  {\bfseries 16} (2024) 076}, \href{http://arxiv.org/abs/2401.09540}{{\ttfamily
  arXiv:2401.09540 [hep-th]}}.

\bibitem{Kapec:2017tkm}
D.~Kapec, M.~Perry, A.-M. Raclariu, and A.~Strominger, ``{Infrared Divergences
  in QED, Revisited},''
  \href{http://dx.doi.org/10.1103/PhysRevD.96.085002}{{\em Phys. Rev.}
  {\bfseries D96} no.~8, (2017) 085002},
\href{http://arxiv.org/abs/1705.04311}{{\ttfamily arXiv:1705.04311 [hep-th]}}.

\bibitem{Chen:2023tvj}
H.~Z. Chen, R.~C. Myers, and A.-M. Raclariu, ``{Entanglement, soft modes, and
  celestial holography},''
  \href{http://dx.doi.org/10.1103/PhysRevD.109.L121702}{{\em Phys. Rev. D}
  {\bfseries 109} no.~12, (2024) L121702},
  \href{http://arxiv.org/abs/2308.12341}{{\ttfamily arXiv:2308.12341
  [hep-th]}}.

\bibitem{Chen:2024kuq}
H.~Z. Chen, R.~Myers, and A.-M. Raclariu, ``{Entanglement, Soft Modes, and
  Celestial CFT},'' \href{http://arxiv.org/abs/2403.13913}{{\ttfamily
  arXiv:2403.13913 [hep-th]}}.

\bibitem{Freidel:2020xyx}
L.~Freidel, M.~Geiller, and D.~Pranzetti, ``{Edge modes of gravity. Part I.
  Corner potentials and charges},''
  \href{http://dx.doi.org/10.1007/JHEP11(2020)026}{{\em JHEP} {\bfseries 11}
  (2020) 026}, \href{http://arxiv.org/abs/2006.12527}{{\ttfamily
  arXiv:2006.12527 [hep-th]}}.

\bibitem{Freidel:2020svx}
L.~Freidel, M.~Geiller, and D.~Pranzetti, ``{Edge modes of gravity. Part II.
  Corner metric and Lorentz charges},''
  \href{http://dx.doi.org/10.1007/JHEP11(2020)027}{{\em JHEP} {\bfseries 11}
  (2020) 027}, \href{http://arxiv.org/abs/2007.03563}{{\ttfamily
  arXiv:2007.03563 [hep-th]}}.

\bibitem{Freidel:2020ayo}
L.~Freidel, M.~Geiller, and D.~Pranzetti, ``{Edge modes of gravity. Part III.
  Corner simplicity constraints},''
  \href{http://dx.doi.org/10.1007/JHEP01(2021)100}{{\em JHEP} {\bfseries 01}
  (2021) 100}, \href{http://arxiv.org/abs/2007.12635}{{\ttfamily
  arXiv:2007.12635 [hep-th]}}.

\bibitem{Freidel:2021cjp}
L.~Freidel, R.~Oliveri, D.~Pranzetti, and S.~Speziale, ``{Extended corner
  symmetry, charge bracket and Einstein's equations},''
  \href{http://dx.doi.org/10.1007/JHEP09(2021)083}{{\em JHEP} {\bfseries 09}
  (2021) 083}, \href{http://arxiv.org/abs/2104.12881}{{\ttfamily
  arXiv:2104.12881 [hep-th]}}.

\bibitem{Wald:1999wa}
R.~M. Wald and A.~Zoupas, ``{A General definition of 'conserved quantities' in
  general relativity and other theories of gravity},''
  \href{http://dx.doi.org/10.1103/PhysRevD.61.084027}{{\em Phys. Rev.}
  {\bfseries D61} (2000) 084027},
\href{http://arxiv.org/abs/gr-qc/9911095}{{\ttfamily arXiv:gr-qc/9911095
  [gr-qc]}}.

\bibitem{Fiorucci:2021pha}
A.~Fiorucci, {\em {Leaky covariant phase spaces: Theory and application to
  $\Lambda$-BMS symmetry}}.
\newblock PhD thesis, Brussels U., Intl. Solvay Inst., Brussels, 2021.
\newblock \href{http://arxiv.org/abs/2112.07666}{{\ttfamily arXiv:2112.07666
  [hep-th]}}.

\bibitem{Mitman:2024uss}
K.~Mitman {\em et al.}, ``{A review of gravitational memory and BMS frame
  fixing in numerical relativity},''
  \href{http://dx.doi.org/10.1088/1361-6382/ad83c2}{{\em Class. Quant. Grav.}
  {\bfseries 41} no.~22, (2024) 223001},
  \href{http://arxiv.org/abs/2405.08868}{{\ttfamily arXiv:2405.08868 [gr-qc]}}.

\bibitem{Aharony:1999ti}
O.~Aharony, S.~S. Gubser, J.~M. Maldacena, H.~Ooguri, and Y.~Oz, ``{Large N
  field theories, string theory and gravity},''
  \href{http://dx.doi.org/10.1016/S0370-1573(99)00083-6}{{\em Phys. Rept.}
  {\bfseries 323} (2000) 183--386},
  \href{http://arxiv.org/abs/hep-th/9905111}{{\ttfamily arXiv:hep-th/9905111}}.

\bibitem{He:2024skc}
T.~He, P.~Mitra, and K.~M. Zurek, ``{A Diamond of Infrared Equivalences in
  Abelian Gauge Theories},'' \href{http://arxiv.org/abs/2405.12303}{{\ttfamily
  arXiv:2405.12303 [hep-th]}}.

\bibitem{Kabat:1992tb}
D.~N. Kabat and M.~Ortiz, ``{Eikonal quantum gravity and Planckian
  scattering},'' \href{http://dx.doi.org/10.1016/0550-3213(92)90627-N}{{\em
  Nucl. Phys. B} {\bfseries 388} (1992) 570--592},
  \href{http://arxiv.org/abs/hep-th/9203082}{{\ttfamily arXiv:hep-th/9203082}}.

\bibitem{Lewkowycz:2013nqa}
A.~Lewkowycz and J.~Maldacena, ``{Generalized gravitational entropy},''
  \href{http://dx.doi.org/10.1007/JHEP08(2013)090}{{\em JHEP} {\bfseries 08}
  (2013) 090}, \href{http://arxiv.org/abs/1304.4926}{{\ttfamily arXiv:1304.4926
  [hep-th]}}.

\bibitem{Pasterski:2017kqt}
S.~Pasterski and S.-H. Shao, ``{Conformal basis for flat space amplitudes},''
  \href{http://dx.doi.org/10.1103/PhysRevD.96.065022}{{\em Phys. Rev. D}
  {\bfseries 96} no.~6, (2017) 065022},
  \href{http://arxiv.org/abs/1705.01027}{{\ttfamily arXiv:1705.01027
  [hep-th]}}.

\bibitem{Raclariu:2021zjz}
A.-M. Raclariu, ``{Lectures on Celestial Holography},''
  \href{http://arxiv.org/abs/2107.02075}{{\ttfamily arXiv:2107.02075
  [hep-th]}}.

\bibitem{Pasterski:2021rjz}
S.~Pasterski, ``{Lectures on celestial amplitudes},''
  \href{http://dx.doi.org/10.1140/epjc/s10052-021-09846-7}{{\em Eur. Phys. J.
  C} {\bfseries 81} no.~12, (2021) 1062},
  \href{http://arxiv.org/abs/2108.04801}{{\ttfamily arXiv:2108.04801
  [hep-th]}}.

\bibitem{Pasterski:2021raf}
S.~Pasterski, M.~Pate, and A.-M. Raclariu, ``{Celestial Holography},'' in {\em
  {Snowmass 2021}}.
\newblock 11, 2021.
\newblock \href{http://arxiv.org/abs/2111.11392}{{\ttfamily arXiv:2111.11392
  [hep-th]}}.

\bibitem{Donnay:2023mrd}
L.~Donnay, ``{Celestial holography: An asymptotic symmetry perspective},''
  \href{http://dx.doi.org/10.1016/j.physrep.2024.04.003}{{\em Phys. Rept.}
  {\bfseries 1073} (2024) 1--41},
  \href{http://arxiv.org/abs/2310.12922}{{\ttfamily arXiv:2310.12922
  [hep-th]}}.

\bibitem{Compere:2018ylh}
G.~Comp\`ere, A.~Fiorucci, and R.~Ruzziconi, ``{Superboost transitions,
  refraction memory and super-Lorentz charge algebra},''
  \href{http://dx.doi.org/10.1007/JHEP11(2018)200}{{\em JHEP} {\bfseries 11}
  (2018) 200}, \href{http://arxiv.org/abs/1810.00377}{{\ttfamily
  arXiv:1810.00377 [hep-th]}}. [Erratum: JHEP 04, 172 (2020)].

\bibitem{Compere:2020lrt}
G.~Comp\`ere, A.~Fiorucci, and R.~Ruzziconi, ``{The $\Lambda$-BMS$_4$ charge
  algebra},'' \href{http://dx.doi.org/10.1007/JHEP10(2020)205}{{\em JHEP}
  {\bfseries 10} (2020) 205}, \href{http://arxiv.org/abs/2004.10769}{{\ttfamily
  arXiv:2004.10769 [hep-th]}}.

\bibitem{Donnay:2021wrk}
L.~Donnay and R.~Ruzziconi, ``{BMS flux algebra in celestial holography},''
  \href{http://dx.doi.org/10.1007/JHEP11(2021)040}{{\em JHEP} {\bfseries 11}
  (2021) 040}, \href{http://arxiv.org/abs/2108.11969}{{\ttfamily
  arXiv:2108.11969 [hep-th]}}.

\bibitem{Kapec:2016aqd}
D.~Kapec, A.-M. Raclariu, and A.~Strominger, ``{Area, Entanglement Entropy and
  Supertranslations at Null Infinity},''
  \href{http://dx.doi.org/10.1088/1361-6382/aa7f12}{{\em Class. Quant. Grav.}
  {\bfseries 34} no.~16, (2017) 165007},
\href{http://arxiv.org/abs/1603.07706}{{\ttfamily arXiv:1603.07706 [hep-th]}}.

\bibitem{Fiorucci:2023lpb}
A.~Fiorucci, D.~Grumiller, and R.~Ruzziconi, ``{Logarithmic celestial conformal
  field theory},'' \href{http://dx.doi.org/10.1103/PhysRevD.109.L021902}{{\em
  Phys. Rev. D} {\bfseries 109} no.~2, (2024) L021902},
  \href{http://arxiv.org/abs/2305.08913}{{\ttfamily arXiv:2305.08913
  [hep-th]}}.

\bibitem{Ciambelli:2024swv}
L.~Ciambelli, L.~Freidel, and R.~G. Leigh, ``{Quantum Null Geometry and
  Gravity},'' \href{http://arxiv.org/abs/2407.11132}{{\ttfamily
  arXiv:2407.11132 [hep-th]}}.

\bibitem{Kapec:2017gsg}
D.~Kapec and P.~Mitra, ``{A $d$-Dimensional Stress Tensor for Mink$_{d+2}$
  Gravity},'' \href{http://dx.doi.org/10.1007/JHEP05(2018)186}{{\em JHEP}
  {\bfseries 05} (2018) 186},
\href{http://arxiv.org/abs/1711.04371}{{\ttfamily arXiv:1711.04371 [hep-th]}}.

\bibitem{Kulish:1970ut}
P.~P. Kulish and L.~D. Faddeev, ``{Asymptotic conditions and infrared
  divergences in quantum electrodynamics},''
  \href{http://dx.doi.org/10.1007/BF01066485}{{\em Theor. Math. Phys.}
  {\bfseries 4} (1970) 745}.

\bibitem{Compere:2018aar}
G.~Comp\`ere and A.~Fiorucci, ``{Advanced Lectures on General Relativity},''
  \href{http://arxiv.org/abs/1801.07064}{{\ttfamily arXiv:1801.07064
  [hep-th]}}.

\bibitem{Harlow:2019yfa}
D.~Harlow and J.-Q. Wu, ``{Covariant phase space with boundaries},''
  \href{http://dx.doi.org/10.1007/JHEP10(2020)146}{{\em JHEP} {\bfseries 10}
  (2020) 146}, \href{http://arxiv.org/abs/1906.08616}{{\ttfamily
  arXiv:1906.08616 [hep-th]}}.

\bibitem{He:2020ifr}
T.~He and P.~Mitra, ``{Covariant Phase Space and Soft Factorization in
  Non-Abelian Gauge Theories},''
  \href{http://dx.doi.org/10.1007/JHEP03(2021)015}{{\em JHEP} {\bfseries 03}
  (2021) 015}, \href{http://arxiv.org/abs/2009.14334}{{\ttfamily
  arXiv:2009.14334 [hep-th]}}.

\end{thebibliography}\endgroup
\bibliographystyle{utphys}

\end{document}